\documentclass[prb,aps,amsmath,amssymb,twocolumn]{revtex4-2}
\usepackage{graphicx,amsmath,amssymb,color,xcolor,orcidlink,bm}
\usepackage{setspace}
\usepackage[utf8]{inputenc}
\usepackage[normalem]{ulem}
\usepackage{nicefrac}
\usepackage{multirow, makecell, ragged2e}
\usepackage[titletoc,title]{appendix}
\usepackage{cancel}
\usepackage[justification=raggedright, singlelinecheck=false]{caption}
\usepackage{adjustbox}
\newcommand{\be}{\begin{equation}}
\newcommand{\ee}{\end{equation}}

\newcommand{\ba}{\begin{eqnarray}}
\newcommand{\ea}{\end{eqnarray}}

\newcommand{\LLL}{\text{P}_{\text{LLL}}}

\def\beq{\begin{eqnarray}}
\def\eeq{\end{eqnarray}}

\newcommand{\QP}[1]{\mathrm{QP}_{#1}}
\newcommand{\QH}[1]{\mathrm{QH}_{#1}}

\newlength{\figwidth}
\begin{document}

\title{Molecular anyons in the fractional quantum Hall effect}
\author{Mytraya Gattu~\orcidlink{0000-0001-6994-389X} and J. K. Jain~\orcidlink{000-0003-0082-5881}}
\affiliation{Department of Physics, 104 Davey Lab, Pennsylvania State University, University Park, Pennsylvania 16802,USA}

\date{\today}
\begin{abstract}
\end{abstract}
\maketitle

{\bf One of the profound consequences of the fractional quantum Hall (FQH) effect~\cite{Tsui82} is the notion of fractionally charged anyons~\cite{Laughlin83,Halperin84,Arovas84}. In spite of extensive experimental study, puzzles remain, however. For example, both shot-noise and Aharonov–Bohm interference measurements sometimes report a charge that is a multiple of the elementary charge~\cite{Heiblum24}. We report here high-precision microscopic calculations that reveal the surprising result that the FQH anyons often bind together into stable clusters, which we term molecular anyons. This is counterintuitive, given that the elementary anyons carry the same charge and are therefore expected to repel one another. The number of anyons in a cluster, its binding energy and its size depend sensitively on the parent FQH state and the interaction between electrons (which is experimentally tunable, e.g., by varying the quantum well width). Our calculations further suggest that the charge-$1/4$ non-Abelian anyons of the $5/2$ FQH state may also bind to form charge-$1/2$ Abelian clusters. The existence of molecular anyons not only can provide a natural explanation for the observed charges, but also leads to a host of new predictions for future experiments and invites a re-analysis of many past ones.}

The FQH effect arises because, at certain fractional fillings of Landau level (LL), such as $\nu=n/(2pn\pm 1)$, inter-electron interactions produce an incompressible liquid with a gap to excitations. Slightly away from these filling factors, the state contains a finite density of quasiparticles (QPs) or quasiholes (QHs). From rather general arguments that do not require a detailed microscopic understanding, it can be shown that these QPs/QHs have fractional charge of magnitude $|e^*|= 1/(2pn\pm 1)$ (in units of the electron charge) and their exchange produces a phase factor $e^{i\theta^*}$ with $\theta^*=\pi [2p/(2pn\pm 1)]$~\cite{Laughlin83,Halperin84}.
Shot noise experiments measured a charge of $e^*\approx 1/3$ at $\nu=1/3$~\cite{dePicciotto97,Saminadayar97}.
At other fractions, such as $\nu=2/5$, $2/3$ and $3/7$, the shot-noise charges have the expected values of $e^*\approx 1/5$, $1/3$ and $1/7$ at some elevated temperatures~\cite{Reznikov99}, but, unexpectedly, the measured charges drift upward as the temperature is lowered, apparently approaching the value $\sim\nu$ in the limit of low temperature~\cite{Chung03,Bid09,Biswas22}.  Explanations have been proposed based on edge physics and upstream neutral modes~\cite{Feldman17,Feldman21,Biswas22} and renormalization group treatment of the relevant tunneling operators~\cite{Ferraro08}. Very recent experiments in chiral Mach-Zehnder geometry have measured the charge at $\nu=2/3,3/5, 4/7$ and found it to be $\sim \nu$ from {\it both} the Aharonov-Bohm (AB) period and shot noise ~\cite{Heiblum24}. Because the edges do not support a sharply quantized charge due to the absence of a gap, we ask if there exists a mechanism for the bulk (the region away from the edge) to facilitate a coherent tunneling of multiple anyons. 

\begin{figure*}
   \textbf{(a)} \begin{adjustbox}{valign=t}
    \begin{minipage}{0.20\textwidth}
        \includegraphics[width=\linewidth]{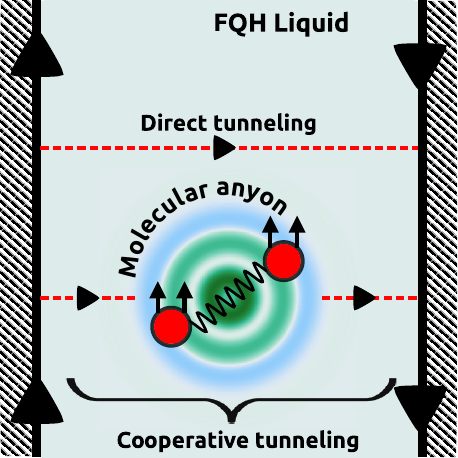}       
    \end{minipage}        
    \end{adjustbox}
    \textbf{(b)}\begin{adjustbox}{valign=t}
        \begin{minipage}{0.60\textwidth}
            \includegraphics[width=0.3\linewidth]{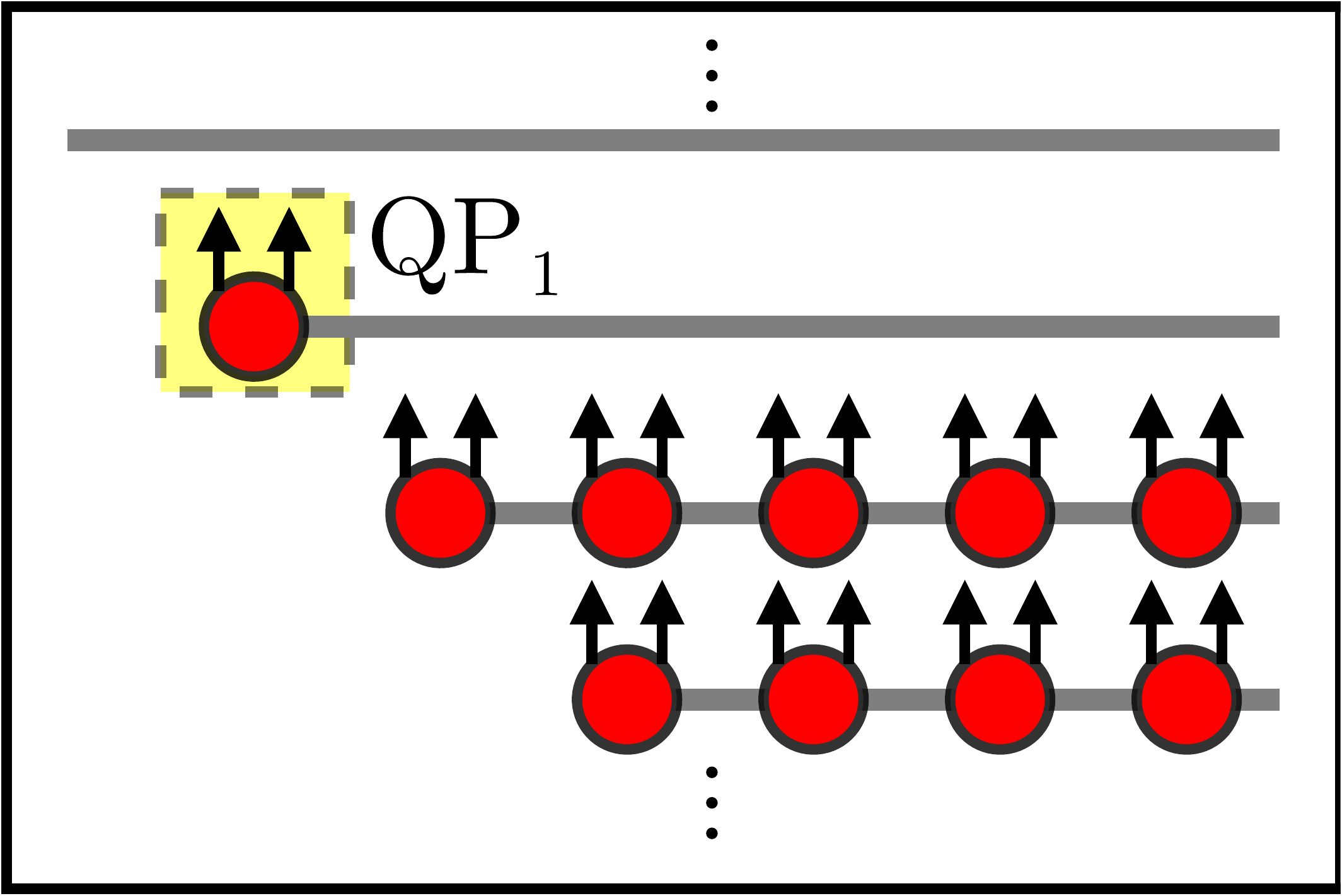}
            \includegraphics[width=0.3\linewidth]{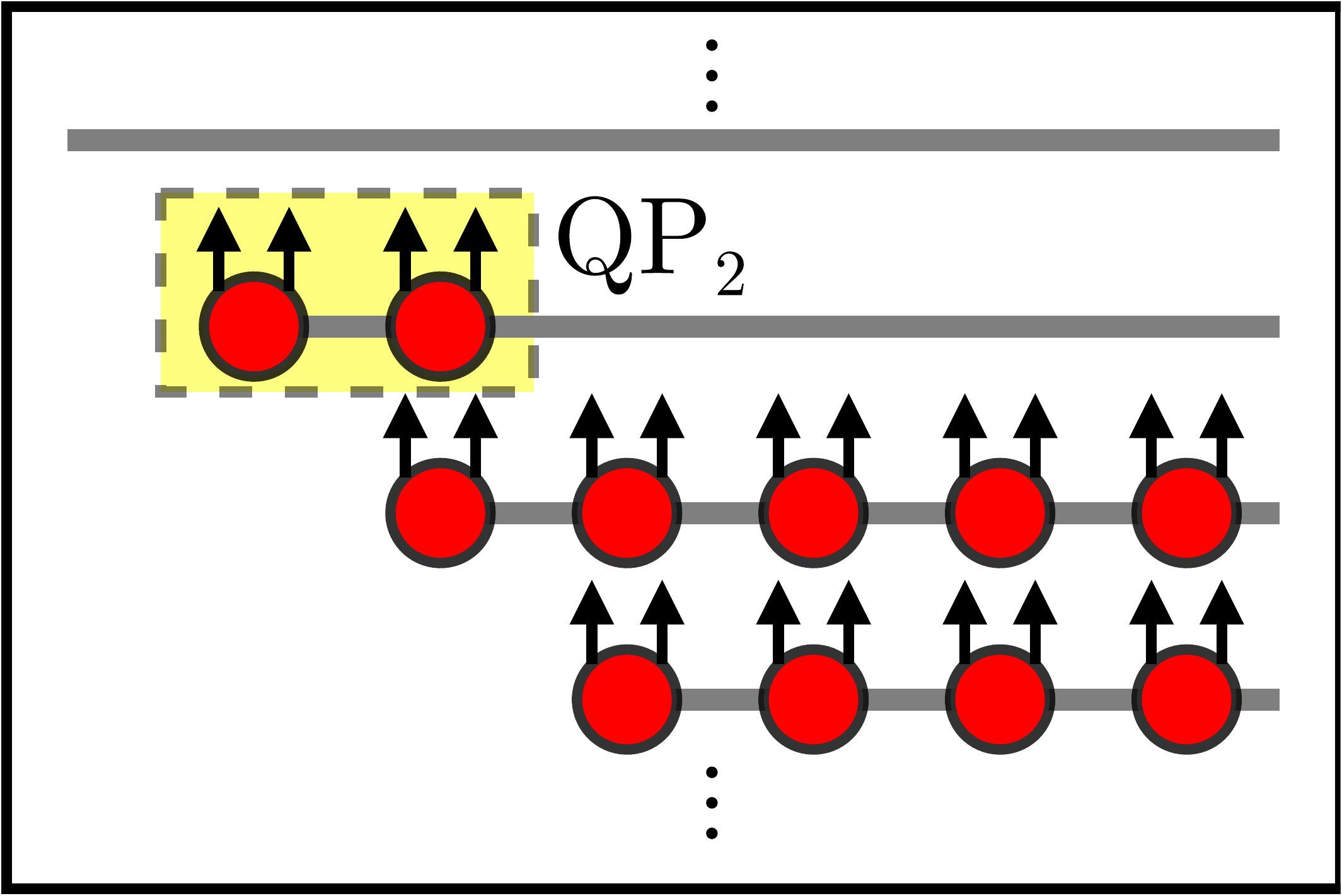}
            \includegraphics[width=0.3\linewidth]{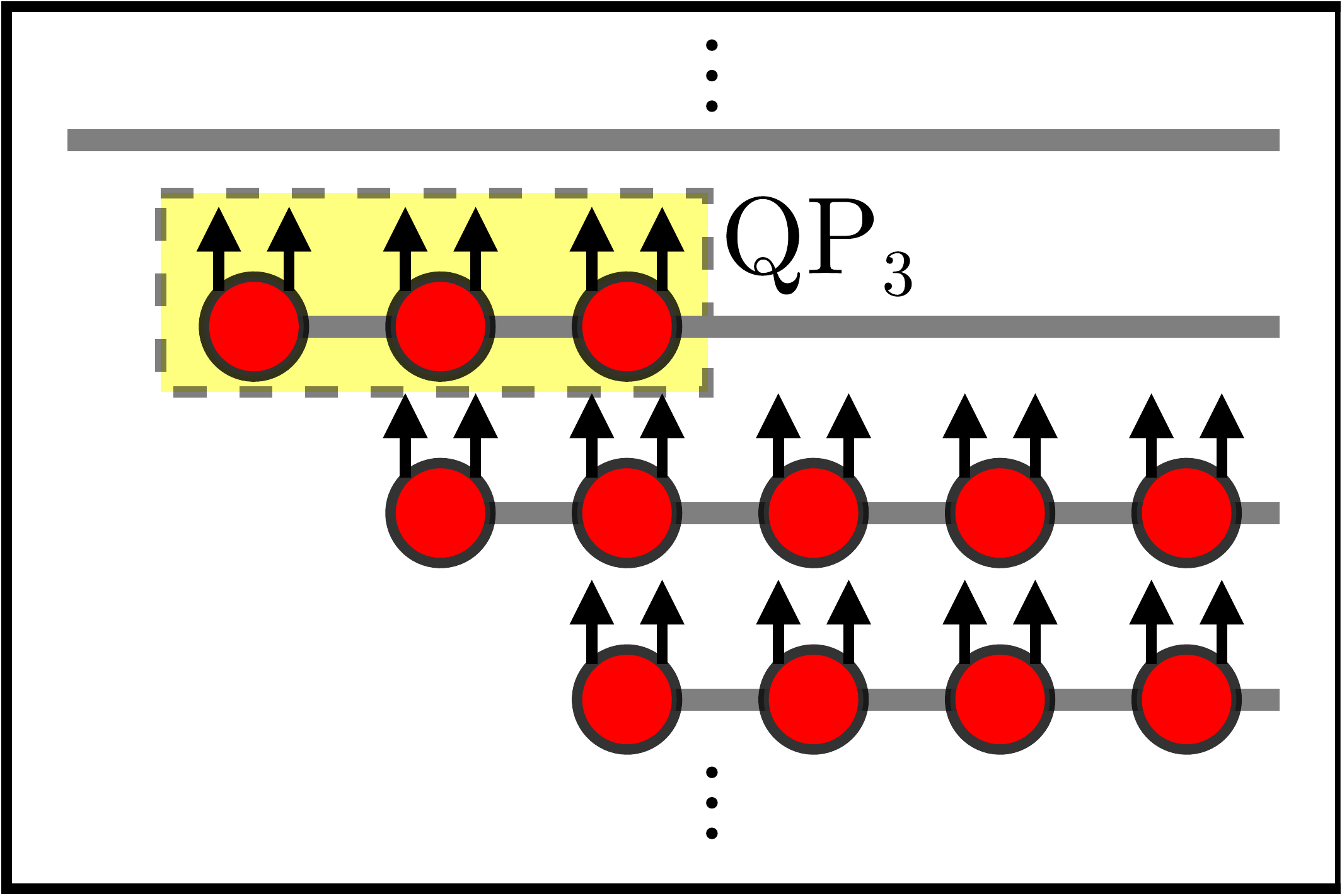}\\
            \includegraphics[width=0.3\linewidth]{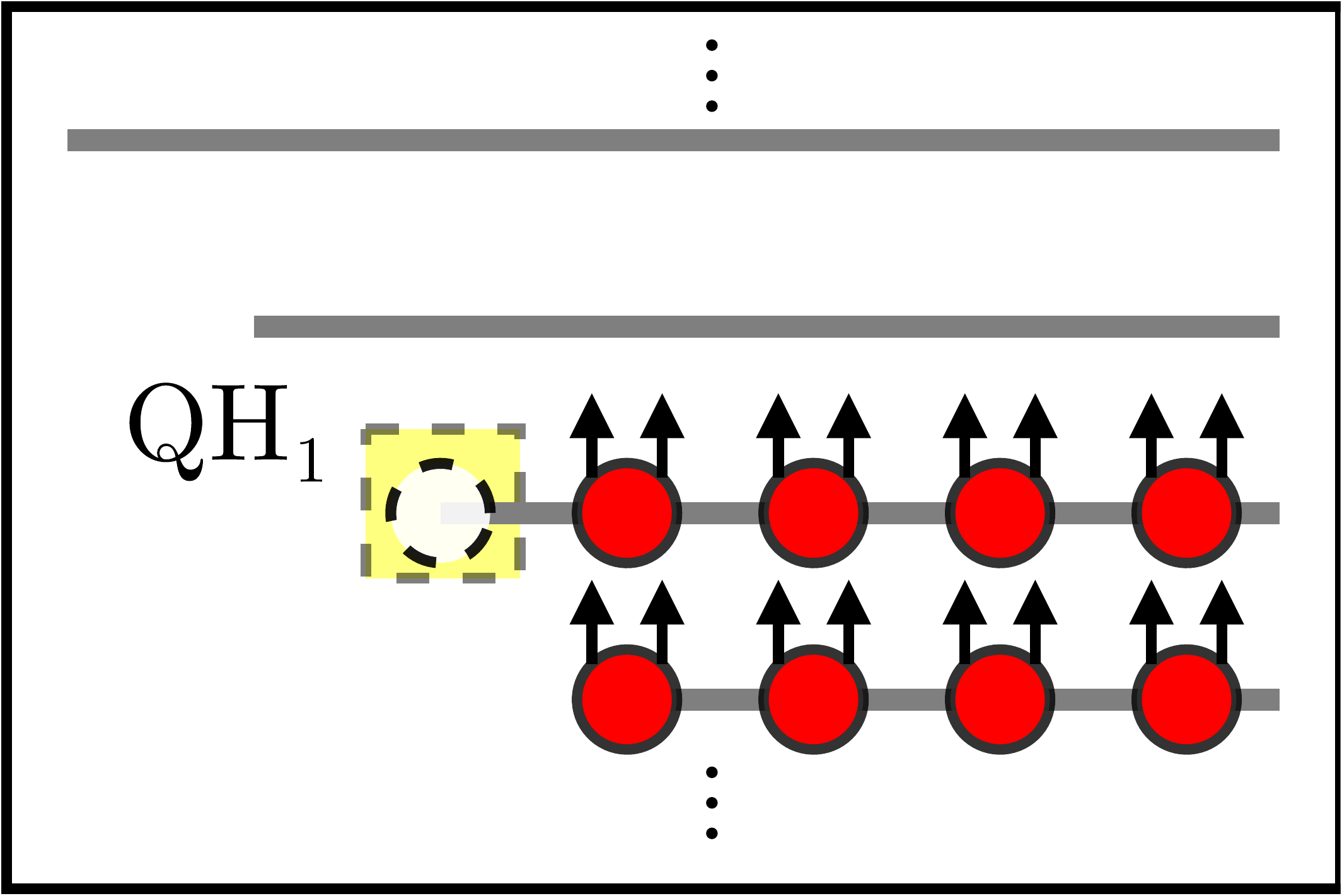}
            \includegraphics[width=0.3\linewidth]{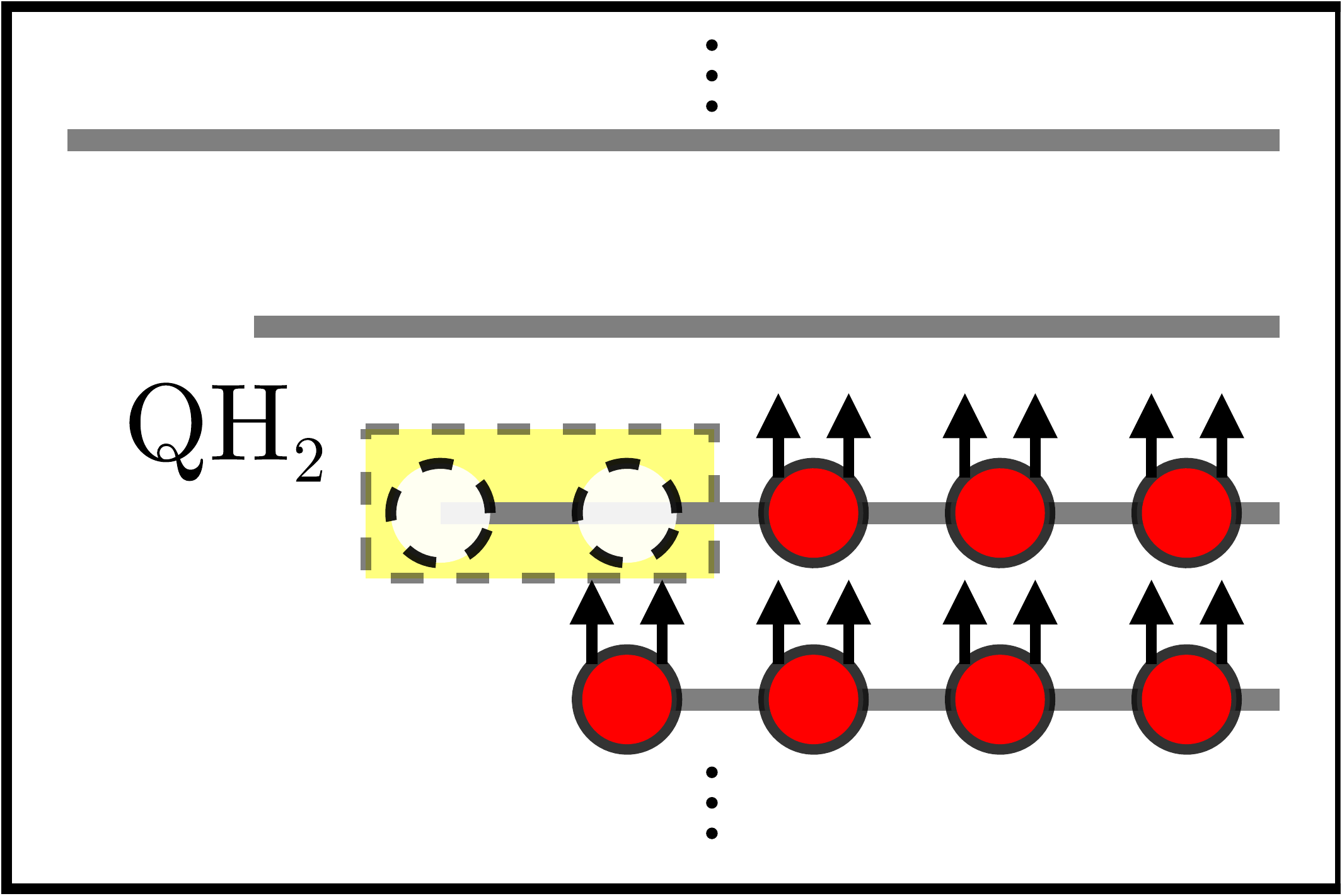}
            \includegraphics[width=0.3\linewidth]{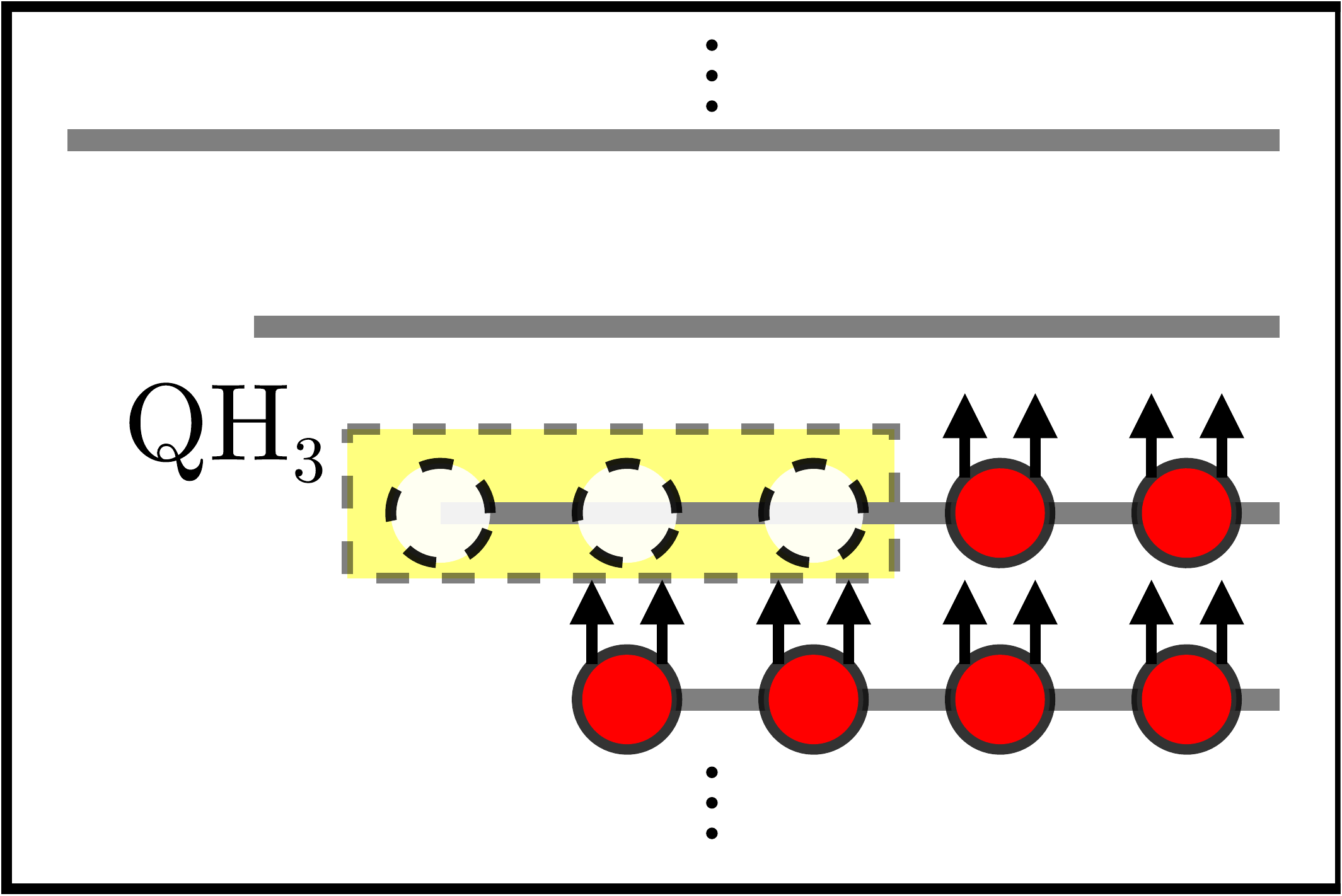}
        \end{minipage}
    \end{adjustbox}
    \newline
    \textbf{(c)} \begin{adjustbox}{valign=t}
        \begin{minipage}{0.87\textwidth}
            \includegraphics[width=\linewidth]{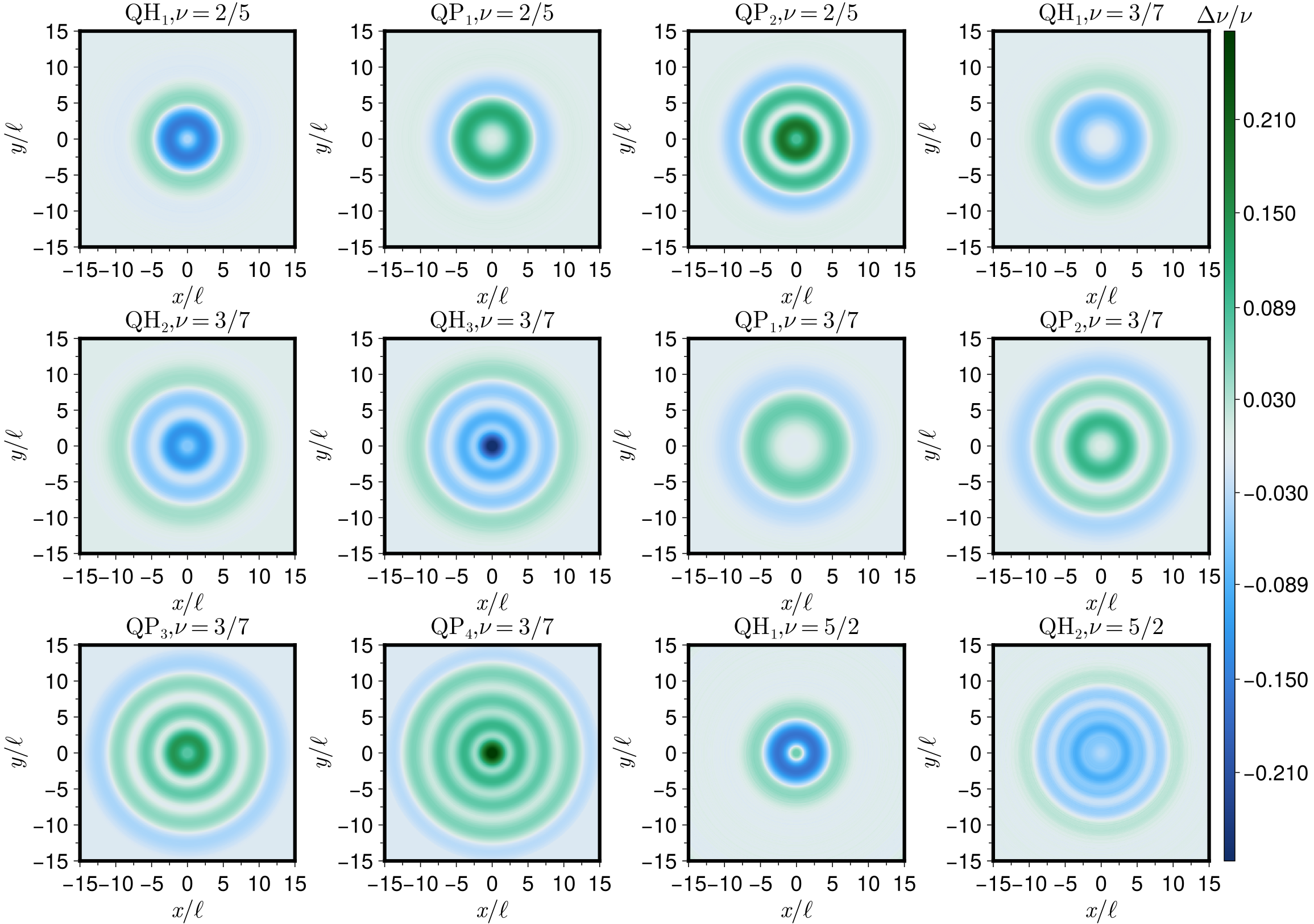}
        \end{minipage}
    \end{adjustbox}
  \caption{(a) Pictorial depiction of direct tunneling from one edge to another (both edges shown as vertical lines) across a constriction, and cooperative tunneling through a localized molecular anyon. The molecular anyon shown here contains two QPs. (b) The panels show the structure of molecular anyons in terms of CF occupation of $\Lambda$ levels; a CF is depicted as an electron with two arrows. $\QP{q}$ consists of an integer number of fully occupied $\Lambda$ levels and additional $q$ CFs in the lowest empty $\Lambda$ level. $\QH{q}$ has an integer number of $\Lambda$ level with $q$ missing CFs from the topmost occupied $\Lambda$ level. (c) The spatial density profiles $\Delta \nu(x,y)/\nu$ of the lowest energy molecular anyons $\QH{q}$ and $\QP{q}$ pinned at the origin for the Jain fractions $\nu=2/5$, $3/7$, and the even-denominator state $\nu=5/2$. The magnetic length is denoted by $\ell$.}
  \label{fig:schematics-densities}
\end{figure*}

The principal result of our work is to demonstrate that the FQH anyons in general bind together to form molecules. We demonstrate this using the precise microscopic description offered by the composite fermion (CF) theory~\cite{Jain89,Jain07}, where the CFs are topological bound states of electrons and an even number of quantum vortices. The strongly interacting electrons in the lowest LL (LLL) capture $2p$ vortices to transform into CFs, which experience a reduced magnetic field $B^{\rm CF}=B-2p\rho\phi_0$ and form Landau-like levels called $\Lambda$ levels ($\Lambda$Ls). For electrons at filling factor $\nu$, CFs fill $\nu^{\rm CF}=\nu/(1-2p\nu)$ $\Lambda$Ls. In particular, for $\nu=n/(2pn+ 1)$, the ground state is $\nu^{\rm CF}=n$ filled $\Lambda$Ls of CFs; a QP is a solitary CF in the $n^{\rm th}$ $\Lambda$L; and a QH is a missing CF in the $(n-1)^{\rm th}$ $\Lambda$ (Fig.~\ref{fig:schematics-densities}b). Wave functions for the ground, QH and QP states at $\nu=n/(2pn+1)$ are constructed as: $\Psi_{n/(2n+1)}=\LLL\Phi_n\Phi_1^{2p}$, $\Psi^{\rm QH}_{n/(2n+1)}=\LLL\Phi^{\rm QH}_n\Phi_1^{2p}$, and $\Psi^{\rm QP}_{n/(2n+1)}=\LLL\Phi^{\rm QP}_n\Phi_1^{2p}$, where $\Phi_n$, $\Phi^{\rm QH}_n$ and $\Phi^{\rm QP}_n$ are the known wave functions of the ground, QP and QH states at $\nu=n$ (see Supplementary Sections 1 and 2 for details). The factor $\Phi_1^{2p}$ attaches $2p$ vortices to electrons and $\LLL$ represents projection into the LLL. These QPs and QHs have quantized fractional charges of magnitude $|e^*|=1/(2pn+ 1)$ and they also obey well-defined braid statistics~\cite{Jeon04}. Furthermore, comparisons with exact diagonalization (ED) studies have shown these to be astonishingly accurate representations of the exact ground states, QPs and QHs for the Coulomb interaction~\cite{Gattu24}. A key advantage of the CF theory is that it allows study large systems~\cite{Kamilla96, Gattu25}, 
which is crucial for the study of molecular anyons, which requires system sizes that can accommodate several well separated QPs or QHs.

We ask whether QPs bind together to form molecules, and if so, what their structure is.  We define the binding energy as $\Delta_q=E^0_{q}-[E^0_{q-1}+E^0_1]$, where $E^0_q$ is the lowest energy bound state of $q$ QPs, determined by the method of CF diagonalization. $\Delta_q$ is the energy change as one QP is brought from infinitely far away and added to QP$_{q-1}$. If $\Delta_q<0$, then the $\QP{q}$ molecule will form. We keep adding QPs one by one until the binding energy becomes positive. The case for QHs is analogous. We find that for the Jain $\nu=n/(2n+1)$ states, the lowest-energy molecules containing $q$ QPs or QHs correspond to the ``compact'' configurations shown in Fig.~\ref{fig:schematics-densities}b, characterized by the minimal relative angular momentum $L_{\mathrm{rel}}=q(q-1)/2$ (see Supplementary Fig. 1 and Supplementary Section 3 for details).

\begin{figure}
    \includegraphics[width=0.9\columnwidth]{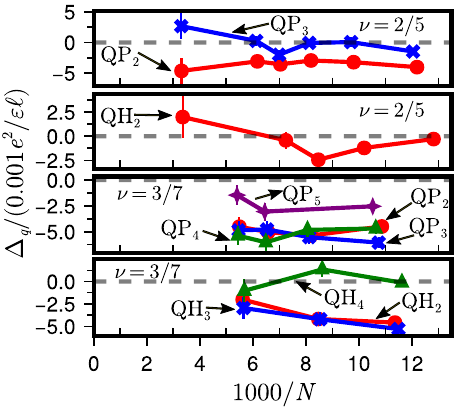}
    \caption{The binding energy $\Delta_q$ (in units of $e^{2}/\varepsilon \ell$) of the molecular anyons $\mathrm{QP}_{q}$ and $\mathrm{QH}_{q}$ at fillings $\nu = 2/5$ and $3/7$ as a function of $1/N$, where $N$ is the number of electrons in the system. The results are for a pure 2D system. A negative intercept in the limit $N^{-1}\rightarrow 0$ indicates stability of the molecule.}
    \label{fig:binding-energies-thermodynamic-limit}
\end{figure}

At $\nu=1/3$, neither QPs nor QHs form bound states~\cite{Lee02}. $\Delta_q$ for several QPs and QHs at $\nu = 2/5$ and $\nu=3/7$ are given in Fig.~\ref{fig:binding-energies-thermodynamic-limit} for a pure two-dimensional system. At $\nu = 2/5$, the thermodynamic binding energies are $\Delta^{\rm QP}_2\sim -0.0025 e^{2}/\varepsilon \ell$ while $\Delta^{\rm QP}_3>0$, implying formation of QP$_2$; in contrast QHs remain unbound.(Here, $\ell=\sqrt{\hbar c/eB}$ is the magnetic length and $\varepsilon$ is the dielectric constant of the background semiconductor.) At $\nu = 3/7$, we find $\Delta^{\rm QP}_2, \Delta^{\rm QP}_3, \Delta^{\rm QP}_4\sim -0.005 e^{2}/\varepsilon \ell$ while $\Delta^{\rm QP}_5>0$. On the QH side, $\Delta^{\rm QH}_{2}, \Delta^{\rm QH}_{3}\sim -0.002 e^{2}/\varepsilon \ell$ while $\Delta^{\rm QH}_{4}>0$. Thus, the system favors the formation of $\mathrm{QP}_{4}$ and QH$_3$ molecules. We note that the binding energies of the molecules are approximately an order of magnitude smaller than the energy required to create a QP or a QH out of the ground state, and thus do not imply an instability of the FQH state.  Given the accuracy of the CF theory, this convincingly demonstrates the formation of molecular anyons in FQH effect. Particle-hole symmetry, which relates $\nu$ to $1-\nu$ for fully polarized states in the absence of LL mixing, implies that $\Delta_q$ for the QH (QP) molecule at $\nu=n/(2n+1)$ is equal to $\Delta_q$ for QP (QH) molecule at $\nu=1-n/(2n+1)$. Fig.~\ref{fig:schematics-densities}c depicts the density profiles of several stable molecular anyons.

Why do anyons not form bound molecules at $\nu=1/3$, and why does the number of anyons, $q$, in the bound molecule go up with increasing $n$ along the Jain sequence $\nu=n/(2n+1)$? The following provides some insight. Because the elementary anyons have the same charge, a Coulomb barrier must be overcome to produce molecular anyons. However, this barrier rapidly goes down with increasing $n$, both because the charge of the elementary anyon $|e^*|=1/(2n+1)$ decreases and the area over which this charge is spread (the size of the anyon) increases.  For instance, as shown in Fig.~\ref{fig:schematics-densities}c, the QPs at $\nu=1/3$, $2/5$ and $3/7$ have characteristic radii of $\sim$ 5$\ell$, $7\ell$ and $8\ell$, encompassing roughly 4-5, 7-9, and 13-14 electrons. While the Coulomb barrier is diminished with increasing $n$, only detailed calculation can tell us what molecules will actually form.

\begin{figure}[t]
    \includegraphics[width=\columnwidth]{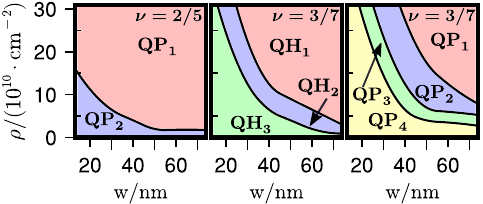}
    \caption{Stability regions for $\QP{q}$ molecules at $\nu = 2/5$ and $\QP{q}$ and $\QH{q}$ molecules at $\nu = 3/7$, plotted as a function of quantum well width $w$ and electron density $\rho$. These results are based on systems with $N \sim 120$, where particles are considered bound if the probability that $\Delta_{q}$ is negative, as estimated from Monte Carlo simulations, exceeds $90\%$. The boundaries are accurate to within $\Delta w = \pm 5\;\mathrm{nm}$ and $\Delta \rho = \pm 1.25  \times 10^{10}\;\mathrm{cm}^{-2}$. At $\nu = 2/5$, only $\QH{1}$ is stabilized.}
  \label{fig:interaction-tuning}
\end{figure}

The nature of the stabilized molecule depends sensitively on the interaction, which can be tuned by increasing the LL mixing (neglected throughout this paper) or by varying the width of the quantum well. Fig.~\ref{fig:interaction-tuning} shows that with increasing the well width $w$ or the electron density $\rho$, molecular anyons with fewer and fewer QPs/QHs are stabilized.

Having studied the odd-denominator fractions $\nu=n/(2n\pm 1)$, we turn to the $\nu = 5/2$ state, where the $\nu=1/2$ state in the second LL is believed to be a topological superconductor of CFs hosting non-Abelian anyons~\cite{Read00}. We address the possibility of the formation of their molecules using the $\nu=1/2$ wave function~\cite{Balram18} 
$\Psi^{\bar{2}\bar{2}111}=\LLL(\Phi_2)^{*2}\Phi_1^3$, belonging to the class of parton wave functions~\cite{Jain89B}. A single QH of this state corresponds to a single QP in $\Phi_2$ (note QP rather than QH, because of the complex conjugation). For two QHs, we have a choice: we can place them either in the same factor or in different factors of $\Phi_2$, which we label as non-topological and topological, respectively (details in Supplementary Section 4). Fig.~\ref{fig:qhs-5-2} shows the binding energy for several relative angular momenta obtained by the method of CF diagonalization for the second LL Coulomb interaction.
 In both cases, the QHs bind to produce $\QH{2}$ molecules. In contrast to the previous cases, the lowest energy molecule is not a compact one but has $L_{\mathrm{rel}} = 5, 6$. The density profiles of the single QH and the topological QH$_2$ are shown in Fig.~\ref{fig:schematics-densities}c. As a note of caution, while $\Psi^{\bar{2}\bar{2}111}$ is reasonably accurate for the ground state~\cite{Balram18}, the wave functions of QPs and QHs do not provide a satisfactory description of the actual wave functions of QPs and QHs obtained in finite system ED studies~\cite{Toke07A}. We are, therefore, not predicting the formation of a QH$_2$ molecule for $\nu=5/2$, but suggesting that as a serious possibility, which is of interest because the $\QH{2}$ is a charge-$1/2$ Abelian object. 

The existence of molecular anyons can be confirmed from direct imaging of the charge density profile of a localized molecule, which has become possible due to recent advances in scanning-tunneling-microscopy (STM) experiments~\cite{Chiu25}. We also expect, in the limit of low disorder, a molecular crystal in the vicinity of $\nu=n/(2n\pm 1)$, which can be distinguished from a QP crystal by the lattice constant, also measurable by STM~\cite{Tsui24}. The predicted asymmetry between the QP and QH molecules should be measurable by studying the behaviors above and below $\nu=n/(2n\pm 1)$. 

We next come to the relevance of molecular anyons to the shot-noise and AB interference measurements of charge and braid statistics. Let us consider the bulk, i.e. the state away from the edges, at zero temperature. When the filling factor is slightly away from $\nu = n/(2n + 1)$, as would invariably be the case in some regions of the constriction, some QPs/QHs are present. These will generically form molecular anyons, which will get pinned by disorder. Now imagine a pinned $\QP{q}$ somewhere near the center of the constriction channel; see Fig.~\ref{fig:schematics-densities}a. In the limit that the temperature is much smaller than the binding energy of $\QP{q}$, a cooperative tunneling of charge $qe^*$ becomes possible: the QP$_q$ tunnels into the edge with lower chemical potential, creating an attractive potential which pulls a charge $qe^*$ from the other edge~\cite{Kivelson24}. This cooperative tunneling of $qe^*$ will dominate over the direct tunneling of a QP (Fig.~\ref{fig:schematics-densities}a), because the tunneling matrix element decays Gaussianly at large distances. This provides a mechanism by which the bulk can filter tunneling of charge $qe^*$, thereby producing a charge $qe^*$ from both shot-noise and AB interference. Note that the charge of the molecule is not necessarily $\nu=ne^*$ (as just happens to be the case at $\nu=1/3$ and $2/5$) but can be higher, e.g., $4/7$ at $\nu=3/7$, or lower as for systems with large quantum well widths and densities. At somewhat higher temperatures, these molecules ionize into elementary charges, which nicely explains, qualitatively, the experimental observation of $e^*$ at an elevated temperature. We note that at $\nu=5/2$ as well, the expected charge $1/4$ is observed at somewhat higher temperatures, which appears to approach $1/2$ at the lowest temperatures~\cite{Dolev08,Dolev10,Ronen25}, consistent with the formation of QH$_2$ molecule.

\begin{figure}
    \includegraphics[width=0.9\columnwidth]{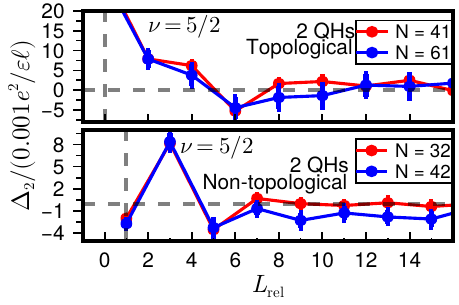}
    \caption{
    Binding energy $\Delta_{2}$ of two QHs of the $\nu=1/2$ non-Abelian parton state $\bar{2}\bar{2}111$, obtained at each relative angular momentum $L_{\mathrm{rel}}$ by CF diagonalization. The energies are measured relative to the state with $L_{\mathrm{rel}} \rightarrow \infty$, which represents far separated QHs. The upper panel considers QHs in distinct parton sectors, i.e. in different $\Phi_2$ factors, referred to as ``topological"; this requires odd $N$. The lower panel considers QHs in the same parton sector, i.e. in the same $\Phi_2$ factor, referred to as ``non-topological"; this corresponds to even $N$.}
    \label{fig:qhs-5-2}
\end{figure}

One may ask whether the Laughlin QH (LQH), which has charge $\nu$, is relevant ~\cite{Heiblum24,Yutushui25A}. An LQH at the origin is described by the wave function $\prod_j z_j\Psi_{n/(2n+1)}$ where $z=x+iy$. In terms of CFs, it is a collection of $n$ QHs, one in each $\Lambda$L. In a mean-field model that treats the CFs as non-interacting, the energy of the LQH is $[n(n-1)/2] \hbar\omega_c^{\rm CF}$ (where $\hbar\omega_c^{\rm CF}$ is the CF cyclotron energy) higher than that of the $\QH{n}$ molecule considered above. Direct calculation shows that the energy of an LQH indeed exceeds that of $\QH{2}$ by $\sim 0.026 e^{2}/\varepsilon \ell$ at $\nu=2/5$, and that of $\QH{3}$ by $\sim 0.05 e^{2}/\varepsilon \ell$ at $\nu=3/7$; these energy differences are significant (an order of magnitude higher than $\Delta_q$), suggesting that the LQH is energetically disfavored for $n>1$. 

The braid statistics in the Fabry-Perot\cite{Nakamura20} and Mach-Zehnder~\cite{Ghosh24, Heiblum24} geometries, manifests through discontinuous slips in the phase associated with a closed loop of a QP as another QP is added or removed from the central island. Our work reveals the possibility that the closed loop could be of a single QP or of QP$_q$, and the object being added could also be a QP or QP$_q$. The phase change of a closed loop of QP$_r$ when a QP$_{q}$ molecule is added in the interior is $2\pi rq [2p/(2pn\pm 1)]$. Thus, if one assumes that depending on details such as the type of disorder and the Coulomb energy, sometimes a single QP might be added and sometimes a molecule, then phase slips of different magnitude are possible. The binding energy can manifest through the period in the gate voltage dependence; for example, for QP$_2$, a smaller change would be required for adding every second QP, thus doubling the period. 

The telegraph noise experiments~\cite{Werkmeister25, Samuelson24} determine all possible curves for the phase. Consider a closed loop of QP$_r$ with QP$_{q}$ molecules being added in the interior one by one. The collection of all possible phase slips, $2\pi rq\{2p/(2pn\pm 1)\}$ (mod $2\pi$), is given by $2\pi \{t/(2pn\pm 1)\}$ with $t=1, 2, \cdots 2pn\pm 1$, provided $r$ and $2pn\pm 1$, and $q$ and $2pn\pm 1$, are relatively prime.  Thus, the collection of measured phase slips as a function of magnetic field for $\nu=1/3$~\cite{Nakamura20} or time ($\nu=1/3, 2/5, 3/7$)~\cite{Werkmeister25, Samuelson24} cannot distinguish whether a QP$_1$ or a $\QP{q}$ molecule is forming a closed loop. However, in some cases it can. An interesting example is $\nu=4/9$: here, QP$_1$ and QP$_2$ produce phase slips of $\{2\pi t/9\}$ with $t= 1,2, \cdots 9$, whereas QP$_3$ will produce $\{2\pi t/3\}$ with $t=1, 2, 3$. QP$_3$ may be stabilized at $\nu=4/9$ by tuning the quantum well width and density.

While the formation of molecular anyons provides insight into certain aspects of the experimental measurements of the charge and statistics, it will take further work for a fully satisfactory understanding. Several experiments see qualitatively different behaviors at $\nu$ and $1-\nu$. For example, the $T=0$ shot-noise charges are in general not the same for $\nu=n/(2n+1)$ and $\nu=1-n/(2n+1)$, and neither are the AB periods seen in chiral Mach-Zehnder geometry~\cite{Ghosh24, Heiblum24}. An understanding of these results would require inclusion of particle-hole symmetry breaking physics. It would also be interesting to explore possible connection with the Chern-Simons Landau-Ginzburg approach that has identified FQH phases analogous to type-I superconductors wherein vortices attract one another leading to a Coulomb frustrated phase separation~\cite{Parameswaran11,Parameswaran12}.

\bibliographystyle{apsrev4-2}
\bibliography{biblio_fqhe_processed.bib}

\begin{thebibliography}{38}%
\makeatletter
\providecommand \@ifxundefined [1]{%
 \@ifx{#1\undefined}
}%
\providecommand \@ifnum [1]{%
 \ifnum #1\expandafter \@firstoftwo
 \else \expandafter \@secondoftwo
 \fi
}%
\providecommand \@ifx [1]{%
 \ifx #1\expandafter \@firstoftwo
 \else \expandafter \@secondoftwo
 \fi
}%
\providecommand \natexlab [1]{#1}%
\providecommand \enquote  [1]{``#1''}%
\providecommand \bibnamefont  [1]{#1}%
\providecommand \bibfnamefont [1]{#1}%
\providecommand \citenamefont [1]{#1}%
\providecommand \href@noop [0]{\@secondoftwo}%
\providecommand \href [0]{\begingroup \@sanitize@url \@href}%
\providecommand \@href[1]{\@@startlink{#1}\@@href}%
\providecommand \@@href[1]{\endgroup#1\@@endlink}%
\providecommand \@sanitize@url [0]{\catcode `\\12\catcode `\$12\catcode
  `\&12\catcode `\#12\catcode `\^12\catcode `\_12\catcode `\%12\relax}%
\providecommand \@@startlink[1]{}%
\providecommand \@@endlink[0]{}%
\providecommand \url  [0]{\begingroup\@sanitize@url \@url }%
\providecommand \@url [1]{\endgroup\@href {#1}{\urlprefix }}%
\providecommand \urlprefix  [0]{URL }%
\providecommand \Eprint [0]{\href }%
\providecommand \doibase [0]{https://doi.org/}%
\providecommand \selectlanguage [0]{\@gobble}%
\providecommand \bibinfo  [0]{\@secondoftwo}%
\providecommand \bibfield  [0]{\@secondoftwo}%
\providecommand \translation [1]{[#1]}%
\providecommand \BibitemOpen [0]{}%
\providecommand \bibitemStop [0]{}%
\providecommand \bibitemNoStop [0]{.\EOS\space}%
\providecommand \EOS [0]{\spacefactor3000\relax}%
\providecommand \BibitemShut  [1]{\csname bibitem#1\endcsname}%
\let\auto@bib@innerbib\@empty
\bibitem [{\citenamefont {Tsui}\ \emph {et~al.}(1982)\citenamefont {Tsui},
  \citenamefont {Stormer},\ and\ \citenamefont {Gossard}}]{Tsui82}%
  \BibitemOpen
  \bibfield  {author} {\bibinfo {author} {\bibfnamefont {D.~C.}\ \bibnamefont
  {Tsui}}, \bibinfo {author} {\bibfnamefont {H.~L.}\ \bibnamefont {Stormer}},\
  and\ \bibinfo {author} {\bibfnamefont {A.~C.}\ \bibnamefont {Gossard}},\
  }\href {https://doi.org/10.1103/PhysRevLett.48.1559} {\bibfield  {journal}
  {\bibinfo  {journal} {Phys. Rev. Lett.}\ }\textbf {\bibinfo {volume} {48}},\
  \bibinfo {pages} {1559} (\bibinfo {year} {1982})}\BibitemShut {NoStop}%
\bibitem [{\citenamefont {Laughlin}(1983)}]{Laughlin83}%
  \BibitemOpen
  \bibfield  {author} {\bibinfo {author} {\bibfnamefont {R.~B.}\ \bibnamefont
  {Laughlin}},\ }\href {https://doi.org/10.1103/PhysRevLett.50.1395} {\bibfield
   {journal} {\bibinfo  {journal} {Phys. Rev. Lett.}\ }\textbf {\bibinfo
  {volume} {50}},\ \bibinfo {pages} {1395} (\bibinfo {year}
  {1983})}\BibitemShut {NoStop}%
\bibitem [{\citenamefont {Halperin}(1984)}]{Halperin84}%
  \BibitemOpen
  \bibfield  {author} {\bibinfo {author} {\bibfnamefont {B.~I.}\ \bibnamefont
  {Halperin}},\ }\href {https://doi.org/10.1103/PhysRevLett.52.1583} {\bibfield
   {journal} {\bibinfo  {journal} {Phys. Rev. Lett.}\ }\textbf {\bibinfo
  {volume} {52}},\ \bibinfo {pages} {1583} (\bibinfo {year}
  {1984})}\BibitemShut {NoStop}%
\bibitem [{\citenamefont {Arovas}\ \emph {et~al.}(1984)\citenamefont {Arovas},
  \citenamefont {Schrieffer},\ and\ \citenamefont {Wilczek}}]{Arovas84}%
  \BibitemOpen
  \bibfield  {author} {\bibinfo {author} {\bibfnamefont {D.}~\bibnamefont
  {Arovas}}, \bibinfo {author} {\bibfnamefont {J.~R.}\ \bibnamefont
  {Schrieffer}},\ and\ \bibinfo {author} {\bibfnamefont {F.}~\bibnamefont
  {Wilczek}},\ }\href {https://doi.org/10.1103/PhysRevLett.53.722} {\bibfield
  {journal} {\bibinfo  {journal} {Phys. Rev. Lett.}\ }\textbf {\bibinfo
  {volume} {53}},\ \bibinfo {pages} {722} (\bibinfo {year} {1984})}\BibitemShut
  {NoStop}%
\bibitem [{\citenamefont {Heiblum}\ \emph {et~al.}(2024)\citenamefont
  {Heiblum}, \citenamefont {Ghosh}, \citenamefont {Labendik}, \citenamefont
  {Umansky},\ and\ \citenamefont {Mross}}]{Heiblum24}%
  \BibitemOpen
  \bibfield  {author} {\bibinfo {author} {\bibfnamefont {M.}~\bibnamefont
  {Heiblum}}, \bibinfo {author} {\bibfnamefont {B.}~\bibnamefont {Ghosh}},
  \bibinfo {author} {\bibfnamefont {M.}~\bibnamefont {Labendik}}, \bibinfo
  {author} {\bibfnamefont {V.}~\bibnamefont {Umansky}},\ and\ \bibinfo {author}
  {\bibfnamefont {D.}~\bibnamefont {Mross}}\ }\href
  {https://doi.org/10.48550/arXiv.2412.16316} {10.48550/arXiv.2412.16316}
  (\bibinfo {year} {2024})\BibitemShut {NoStop}%
\bibitem [{\citenamefont {de~Picciotto}\ \emph {et~al.}(1997)\citenamefont
  {de~Picciotto}, \citenamefont {Reznikov}, \citenamefont {Heiblum},
  \citenamefont {Umansky}, \citenamefont {Bunin},\ and\ \citenamefont
  {Mahalu}}]{dePicciotto97}%
  \BibitemOpen
  \bibfield  {author} {\bibinfo {author} {\bibfnamefont {R.}~\bibnamefont
  {de~Picciotto}}, \bibinfo {author} {\bibfnamefont {M.}~\bibnamefont
  {Reznikov}}, \bibinfo {author} {\bibfnamefont {M.}~\bibnamefont {Heiblum}},
  \bibinfo {author} {\bibfnamefont {V.}~\bibnamefont {Umansky}}, \bibinfo
  {author} {\bibfnamefont {G.}~\bibnamefont {Bunin}},\ and\ \bibinfo {author}
  {\bibfnamefont {D.}~\bibnamefont {Mahalu}},\ }\href@noop {} {\bibfield
  {journal} {\bibinfo  {journal} {Nature}\ }\textbf {\bibinfo {volume} {389}},\
  \bibinfo {pages} {162} (\bibinfo {year} {1997})}\BibitemShut {NoStop}%
\bibitem [{\citenamefont {Saminadayar}\ \emph {et~al.}(1997)\citenamefont
  {Saminadayar}, \citenamefont {Glattli}, \citenamefont {Jin},\ and\
  \citenamefont {Etienne}}]{Saminadayar97}%
  \BibitemOpen
  \bibfield  {author} {\bibinfo {author} {\bibfnamefont {L.}~\bibnamefont
  {Saminadayar}}, \bibinfo {author} {\bibfnamefont {D.~C.}\ \bibnamefont
  {Glattli}}, \bibinfo {author} {\bibfnamefont {Y.}~\bibnamefont {Jin}},\ and\
  \bibinfo {author} {\bibfnamefont {B.}~\bibnamefont {Etienne}},\ }\href
  {https://doi.org/10.1103/PhysRevLett.79.2526} {\bibfield  {journal} {\bibinfo
   {journal} {Phys. Rev. Lett.}\ }\textbf {\bibinfo {volume} {79}},\ \bibinfo
  {pages} {2526} (\bibinfo {year} {1997})}\BibitemShut {NoStop}%
\bibitem [{\citenamefont {Reznikov}\ \emph {et~al.}(1999)\citenamefont
  {Reznikov}, \citenamefont {Picciotto}, \citenamefont {Griffiths},
  \citenamefont {Heiblum},\ and\ \citenamefont {Umansky}}]{Reznikov99}%
  \BibitemOpen
  \bibfield  {author} {\bibinfo {author} {\bibfnamefont {M.}~\bibnamefont
  {Reznikov}}, \bibinfo {author} {\bibfnamefont {R.~d.}\ \bibnamefont
  {Picciotto}}, \bibinfo {author} {\bibfnamefont {T.}~\bibnamefont
  {Griffiths}}, \bibinfo {author} {\bibfnamefont {M.}~\bibnamefont {Heiblum}},\
  and\ \bibinfo {author} {\bibfnamefont {V.}~\bibnamefont {Umansky}},\
  }\href@noop {} {\bibfield  {journal} {\bibinfo  {journal} {Nature}\ }\textbf
  {\bibinfo {volume} {399}},\ \bibinfo {pages} {238} (\bibinfo {year}
  {1999})}\BibitemShut {NoStop}%
\bibitem [{\citenamefont {Chung}\ \emph {et~al.}(2003)\citenamefont {Chung},
  \citenamefont {Heiblum},\ and\ \citenamefont {Umansky}}]{Chung03}%
  \BibitemOpen
  \bibfield  {author} {\bibinfo {author} {\bibfnamefont {Y.~C.}\ \bibnamefont
  {Chung}}, \bibinfo {author} {\bibfnamefont {M.}~\bibnamefont {Heiblum}},\
  and\ \bibinfo {author} {\bibfnamefont {V.}~\bibnamefont {Umansky}},\ }\href
  {https://doi.org/10.1103/PhysRevLett.91.216804} {\bibfield  {journal}
  {\bibinfo  {journal} {Phys. Rev. Lett.}\ }\textbf {\bibinfo {volume} {91}},\
  \bibinfo {pages} {216804} (\bibinfo {year} {2003})}\BibitemShut {NoStop}%
\bibitem [{\citenamefont {Bid}\ \emph {et~al.}(2009)\citenamefont {Bid},
  \citenamefont {Ofek}, \citenamefont {Heiblum}, \citenamefont {Umansky},\ and\
  \citenamefont {Mahalu}}]{Bid09}%
  \BibitemOpen
  \bibfield  {author} {\bibinfo {author} {\bibfnamefont {A.}~\bibnamefont
  {Bid}}, \bibinfo {author} {\bibfnamefont {N.}~\bibnamefont {Ofek}}, \bibinfo
  {author} {\bibfnamefont {M.}~\bibnamefont {Heiblum}}, \bibinfo {author}
  {\bibfnamefont {V.}~\bibnamefont {Umansky}},\ and\ \bibinfo {author}
  {\bibfnamefont {D.}~\bibnamefont {Mahalu}},\ }\href
  {https://doi.org/10.1103/PhysRevLett.103.236802} {\bibfield  {journal}
  {\bibinfo  {journal} {Phys. Rev. Lett.}\ }\textbf {\bibinfo {volume} {103}},\
  \bibinfo {pages} {236802} (\bibinfo {year} {2009})}\BibitemShut {NoStop}%
\bibitem [{\citenamefont {Biswas}\ \emph {et~al.}(2022)\citenamefont {Biswas},
  \citenamefont {Bhattacharyya}, \citenamefont {Kundu}, \citenamefont {Das},
  \citenamefont {Heiblum}, \citenamefont {Umansky}, \citenamefont {Goldstein},\
  and\ \citenamefont {Gefen}}]{Biswas22}%
  \BibitemOpen
  \bibfield  {author} {\bibinfo {author} {\bibfnamefont {S.}~\bibnamefont
  {Biswas}}, \bibinfo {author} {\bibfnamefont {R.}~\bibnamefont
  {Bhattacharyya}}, \bibinfo {author} {\bibfnamefont {H.~K.}\ \bibnamefont
  {Kundu}}, \bibinfo {author} {\bibfnamefont {A.}~\bibnamefont {Das}}, \bibinfo
  {author} {\bibfnamefont {M.}~\bibnamefont {Heiblum}}, \bibinfo {author}
  {\bibfnamefont {V.}~\bibnamefont {Umansky}}, \bibinfo {author} {\bibfnamefont
  {M.}~\bibnamefont {Goldstein}},\ and\ \bibinfo {author} {\bibfnamefont
  {Y.}~\bibnamefont {Gefen}},\ }\href@noop {} {\bibfield  {journal} {\bibinfo
  {journal} {Nature physics}\ }\textbf {\bibinfo {volume} {18}},\ \bibinfo
  {pages} {1476} (\bibinfo {year} {2022})}\BibitemShut {NoStop}%
\bibitem [{\citenamefont {Feldman}\ and\ \citenamefont
  {Heiblum}(2017)}]{Feldman17}%
  \BibitemOpen
  \bibfield  {author} {\bibinfo {author} {\bibfnamefont {D.~E.}\ \bibnamefont
  {Feldman}}\ and\ \bibinfo {author} {\bibfnamefont {M.}~\bibnamefont
  {Heiblum}},\ }\href {https://doi.org/10.1103/PhysRevB.95.115308} {\bibfield
  {journal} {\bibinfo  {journal} {Phys. Rev. B}\ }\textbf {\bibinfo {volume}
  {95}},\ \bibinfo {pages} {115308} (\bibinfo {year} {2017})}\BibitemShut
  {NoStop}%
\bibitem [{\citenamefont {Feldman}\ and\ \citenamefont
  {Halperin}(2021)}]{Feldman21}%
  \BibitemOpen
  \bibfield  {author} {\bibinfo {author} {\bibfnamefont {D.~E.}\ \bibnamefont
  {Feldman}}\ and\ \bibinfo {author} {\bibfnamefont {B.~I.}\ \bibnamefont
  {Halperin}},\ }\href@noop {} {\bibfield  {journal} {\bibinfo  {journal}
  {Reports on Progress in Physics}\ }\textbf {\bibinfo {volume} {84}},\
  \bibinfo {pages} {076501} (\bibinfo {year} {2021})}\BibitemShut {NoStop}%
\bibitem [{\citenamefont {Ferraro}\ \emph {et~al.}(2008)\citenamefont
  {Ferraro}, \citenamefont {Braggio}, \citenamefont {Merlo}, \citenamefont
  {Magnoli},\ and\ \citenamefont {Sassetti}}]{Ferraro08}%
  \BibitemOpen
  \bibfield  {author} {\bibinfo {author} {\bibfnamefont {D.}~\bibnamefont
  {Ferraro}}, \bibinfo {author} {\bibfnamefont {A.}~\bibnamefont {Braggio}},
  \bibinfo {author} {\bibfnamefont {M.}~\bibnamefont {Merlo}}, \bibinfo
  {author} {\bibfnamefont {N.}~\bibnamefont {Magnoli}},\ and\ \bibinfo {author}
  {\bibfnamefont {M.}~\bibnamefont {Sassetti}},\ }\href
  {https://doi.org/10.1103/PhysRevLett.101.166805} {\bibfield  {journal}
  {\bibinfo  {journal} {Phys. Rev. Lett.}\ }\textbf {\bibinfo {volume} {101}},\
  \bibinfo {pages} {166805} (\bibinfo {year} {2008})}\BibitemShut {NoStop}%
\bibitem [{\citenamefont {Jain}(1989{\natexlab{a}})}]{Jain89}%
  \BibitemOpen
  \bibfield  {author} {\bibinfo {author} {\bibfnamefont {J.~K.}\ \bibnamefont
  {Jain}},\ }\href {https://doi.org/10.1103/PhysRevLett.63.199} {\bibfield
  {journal} {\bibinfo  {journal} {Phys. Rev. Lett.}\ }\textbf {\bibinfo
  {volume} {63}},\ \bibinfo {pages} {199} (\bibinfo {year}
  {1989}{\natexlab{a}})}\BibitemShut {NoStop}%
\bibitem [{\citenamefont {Jain}(2007)}]{Jain07}%
  \BibitemOpen
  \bibfield  {author} {\bibinfo {author} {\bibfnamefont {J.~K.}\ \bibnamefont
  {Jain}},\ }\href@noop {} {\emph {\bibinfo {title} {Composite Fermions}}}\
  (\bibinfo  {publisher} {Cambridge University Press, New York, US},\ \bibinfo
  {year} {2007})\BibitemShut {NoStop}%
\bibitem [{\citenamefont {Jeon}\ \emph {et~al.}(2004)\citenamefont {Jeon},
  \citenamefont {Graham},\ and\ \citenamefont {Jain}}]{Jeon04}%
  \BibitemOpen
  \bibfield  {author} {\bibinfo {author} {\bibfnamefont {G.~S.}\ \bibnamefont
  {Jeon}}, \bibinfo {author} {\bibfnamefont {K.~L.}\ \bibnamefont {Graham}},\
  and\ \bibinfo {author} {\bibfnamefont {J.~K.}\ \bibnamefont {Jain}},\ }\href
  {https://doi.org/10.1103/PhysRevB.70.125316} {\bibfield  {journal} {\bibinfo
  {journal} {Phys. Rev. B}\ }\textbf {\bibinfo {volume} {70}},\ \bibinfo
  {pages} {125316} (\bibinfo {year} {2004})}\BibitemShut {NoStop}%
\bibitem [{\citenamefont {Gattu}\ \emph {et~al.}(2024)\citenamefont {Gattu},
  \citenamefont {Sreejith},\ and\ \citenamefont {Jain}}]{Gattu24}%
  \BibitemOpen
  \bibfield  {author} {\bibinfo {author} {\bibfnamefont {M.}~\bibnamefont
  {Gattu}}, \bibinfo {author} {\bibfnamefont {G.~J.}\ \bibnamefont
  {Sreejith}},\ and\ \bibinfo {author} {\bibfnamefont {J.~K.}\ \bibnamefont
  {Jain}},\ }\href {https://doi.org/10.1103/PhysRevB.109.L201123} {\bibfield
  {journal} {\bibinfo  {journal} {Phys. Rev. B}\ }\textbf {\bibinfo {volume}
  {109}},\ \bibinfo {pages} {L201123} (\bibinfo {year} {2024})}\BibitemShut
  {NoStop}%
\bibitem [{\citenamefont {Kamilla}\ \emph {et~al.}(1996)\citenamefont
  {Kamilla}, \citenamefont {Wu},\ and\ \citenamefont {Jain}}]{Kamilla96}%
  \BibitemOpen
  \bibfield  {author} {\bibinfo {author} {\bibfnamefont {R.~K.}\ \bibnamefont
  {Kamilla}}, \bibinfo {author} {\bibfnamefont {X.~G.}\ \bibnamefont {Wu}},\
  and\ \bibinfo {author} {\bibfnamefont {J.~K.}\ \bibnamefont {Jain}},\
  }\href@noop {} {\bibfield  {journal} {\bibinfo  {journal} {Solid State
  Commun.}\ }\textbf {\bibinfo {volume} {99}} (\bibinfo {year}
  {1996})}\BibitemShut {NoStop}%
\bibitem [{\citenamefont {Gattu}\ and\ \citenamefont {Jain}(2025)}]{Gattu25}%
  \BibitemOpen
  \bibfield  {author} {\bibinfo {author} {\bibfnamefont {M.}~\bibnamefont
  {Gattu}}\ and\ \bibinfo {author} {\bibfnamefont {J.}~\bibnamefont {Jain}},\
  }\bibfield  {journal} {\bibinfo  {journal} {Physical Review Letters}\
  }\textbf {\bibinfo {volume} {134}},\ \href
  {https://doi.org/10.1103/physrevlett.134.156501}
  {10.1103/physrevlett.134.156501} (\bibinfo {year} {2025})\BibitemShut
  {NoStop}%
\bibitem [{\citenamefont {Lee}\ \emph {et~al.}(2002)\citenamefont {Lee},
  \citenamefont {Scarola},\ and\ \citenamefont {Jain}}]{Lee02}%
  \BibitemOpen
  \bibfield  {author} {\bibinfo {author} {\bibfnamefont {S.-Y.}\ \bibnamefont
  {Lee}}, \bibinfo {author} {\bibfnamefont {V.~W.}\ \bibnamefont {Scarola}},\
  and\ \bibinfo {author} {\bibfnamefont {J.~K.}\ \bibnamefont {Jain}},\ }\href
  {https://doi.org/10.1103/PhysRevB.66.085336} {\bibfield  {journal} {\bibinfo
  {journal} {Phys. Rev. B}\ }\textbf {\bibinfo {volume} {66}},\ \bibinfo
  {pages} {085336} (\bibinfo {year} {2002})}\BibitemShut {NoStop}%
\bibitem [{\citenamefont {Read}\ and\ \citenamefont {Green}(2000)}]{Read00}%
  \BibitemOpen
  \bibfield  {author} {\bibinfo {author} {\bibfnamefont {N.}~\bibnamefont
  {Read}}\ and\ \bibinfo {author} {\bibfnamefont {D.}~\bibnamefont {Green}},\
  }\href {https://doi.org/10.1103/PhysRevB.61.10267} {\bibfield  {journal}
  {\bibinfo  {journal} {Phys. Rev. B}\ }\textbf {\bibinfo {volume} {61}},\
  \bibinfo {pages} {10267} (\bibinfo {year} {2000})}\BibitemShut {NoStop}%
\bibitem [{\citenamefont {Balram}\ \emph {et~al.}(2018)\citenamefont {Balram},
  \citenamefont {Barkeshli},\ and\ \citenamefont {Rudner}}]{Balram18}%
  \BibitemOpen
  \bibfield  {author} {\bibinfo {author} {\bibfnamefont {A.~C.}\ \bibnamefont
  {Balram}}, \bibinfo {author} {\bibfnamefont {M.}~\bibnamefont {Barkeshli}},\
  and\ \bibinfo {author} {\bibfnamefont {M.~S.}\ \bibnamefont {Rudner}},\
  }\href {https://doi.org/10.1103/PhysRevB.98.035127} {\bibfield  {journal}
  {\bibinfo  {journal} {Phys. Rev. B}\ }\textbf {\bibinfo {volume} {98}},\
  \bibinfo {pages} {035127} (\bibinfo {year} {2018})}\BibitemShut {NoStop}%
\bibitem [{\citenamefont {Jain}(1989{\natexlab{b}})}]{Jain89B}%
  \BibitemOpen
  \bibfield  {author} {\bibinfo {author} {\bibfnamefont {J.~K.}\ \bibnamefont
  {Jain}},\ }\href {https://doi.org/10.1103/PhysRevB.40.8079} {\bibfield
  {journal} {\bibinfo  {journal} {Phys. Rev. B}\ }\textbf {\bibinfo {volume}
  {40}},\ \bibinfo {pages} {8079} (\bibinfo {year}
  {1989}{\natexlab{b}})}\BibitemShut {NoStop}%
\bibitem [{\citenamefont {T\ifmmode~\mbox{\H{o}}\else \H{o}\fi{}ke}\ \emph
  {et~al.}(2007)\citenamefont {T\ifmmode~\mbox{\H{o}}\else \H{o}\fi{}ke},
  \citenamefont {Regnault},\ and\ \citenamefont {Jain}}]{Toke07A}%
  \BibitemOpen
  \bibfield  {author} {\bibinfo {author} {\bibfnamefont {C.}~\bibnamefont
  {T\ifmmode~\mbox{\H{o}}\else \H{o}\fi{}ke}}, \bibinfo {author} {\bibfnamefont
  {N.}~\bibnamefont {Regnault}},\ and\ \bibinfo {author} {\bibfnamefont
  {J.~K.}\ \bibnamefont {Jain}},\ }\href
  {https://doi.org/10.1103/PhysRevLett.98.036806} {\bibfield  {journal}
  {\bibinfo  {journal} {Phys. Rev. Lett.}\ }\textbf {\bibinfo {volume} {98}},\
  \bibinfo {pages} {036806} (\bibinfo {year} {2007})}\BibitemShut {NoStop}%
\bibitem [{\citenamefont {Chiu}\ \emph {et~al.}(2025)\citenamefont {Chiu},
  \citenamefont {Wang}, \citenamefont {Fan}, \citenamefont {Watanabe},
  \citenamefont {Taniguchi}, \citenamefont {Liu}, \citenamefont {Zaletel},\
  and\ \citenamefont {Yazdani}}]{Chiu25}%
  \BibitemOpen
  \bibfield  {author} {\bibinfo {author} {\bibfnamefont {C.-L.}\ \bibnamefont
  {Chiu}}, \bibinfo {author} {\bibfnamefont {T.}~\bibnamefont {Wang}}, \bibinfo
  {author} {\bibfnamefont {R.}~\bibnamefont {Fan}}, \bibinfo {author}
  {\bibfnamefont {K.}~\bibnamefont {Watanabe}}, \bibinfo {author}
  {\bibfnamefont {T.}~\bibnamefont {Taniguchi}}, \bibinfo {author}
  {\bibfnamefont {X.}~\bibnamefont {Liu}}, \bibinfo {author} {\bibfnamefont
  {M.~P.}\ \bibnamefont {Zaletel}},\ and\ \bibinfo {author} {\bibfnamefont
  {A.}~\bibnamefont {Yazdani}},\ }\bibfield  {journal} {\bibinfo  {journal}
  {Proceedings of the National Academy of Sciences}\ }\textbf {\bibinfo
  {volume} {122}},\ \href {https://doi.org/10.1073/pnas.2424781122}
  {10.1073/pnas.2424781122} (\bibinfo {year} {2025})\BibitemShut {NoStop}%
\bibitem [{\citenamefont {Tsui}\ \emph {et~al.}(2024)\citenamefont {Tsui},
  \citenamefont {He}, \citenamefont {Hu}, \citenamefont {Lake}, \citenamefont
  {Wang}, \citenamefont {Watanabe}, \citenamefont {Taniguchi}, \citenamefont
  {Zaletel},\ and\ \citenamefont {Yazdani}}]{Tsui24}%
  \BibitemOpen
  \bibfield  {author} {\bibinfo {author} {\bibfnamefont {Y.-C.}\ \bibnamefont
  {Tsui}}, \bibinfo {author} {\bibfnamefont {M.}~\bibnamefont {He}}, \bibinfo
  {author} {\bibfnamefont {Y.}~\bibnamefont {Hu}}, \bibinfo {author}
  {\bibfnamefont {E.}~\bibnamefont {Lake}}, \bibinfo {author} {\bibfnamefont
  {T.}~\bibnamefont {Wang}}, \bibinfo {author} {\bibfnamefont {K.}~\bibnamefont
  {Watanabe}}, \bibinfo {author} {\bibfnamefont {T.}~\bibnamefont {Taniguchi}},
  \bibinfo {author} {\bibfnamefont {M.~P.}\ \bibnamefont {Zaletel}},\ and\
  \bibinfo {author} {\bibfnamefont {A.}~\bibnamefont {Yazdani}},\ }\href
  {https://doi.org/10.1038/s41586-024-07212-7} {\bibfield  {journal} {\bibinfo
  {journal} {Nature}\ }\textbf {\bibinfo {volume} {628}},\ \bibinfo {pages}
  {287–292} (\bibinfo {year} {2024})}\BibitemShut {NoStop}%
\bibitem [{\citenamefont {Kivelson}\ and\ \citenamefont
  {Murthy}(2024)}]{Kivelson24}%
  \BibitemOpen
  \bibfield  {author} {\bibinfo {author} {\bibfnamefont {S.~A.}\ \bibnamefont
  {Kivelson}}\ and\ \bibinfo {author} {\bibfnamefont {C.}~\bibnamefont
  {Murthy}},\ }\href {https://doi.org/10.48550/ARXIV.2403.12139} {\bibinfo
  {title} {A modified interferometer to measure anyonic braiding statistics}}
  (\bibinfo {year} {2024})\BibitemShut {NoStop}%
\bibitem [{\citenamefont {Dolev}\ \emph {et~al.}(2008)\citenamefont {Dolev},
  \citenamefont {Heiblum}, \citenamefont {Stern}, \citenamefont {Umansky},\
  and\ \citenamefont {Mahalu}}]{Dolev08}%
  \BibitemOpen
  \bibfield  {author} {\bibinfo {author} {\bibfnamefont {M.}~\bibnamefont
  {Dolev}}, \bibinfo {author} {\bibfnamefont {M.}~\bibnamefont {Heiblum}},
  \bibinfo {author} {\bibfnamefont {A.}~\bibnamefont {Stern}}, \bibinfo
  {author} {\bibfnamefont {V.}~\bibnamefont {Umansky}},\ and\ \bibinfo {author}
  {\bibfnamefont {D.}~\bibnamefont {Mahalu}},\ }\href
  {https://doi.org/10.1038/nature06855} {\bibfield  {journal} {\bibinfo
  {journal} {Nature}\ }\textbf {\bibinfo {volume} {452}},\ \bibinfo {pages}
  {829 EP } (\bibinfo {year} {2008})}\BibitemShut {NoStop}%
\bibitem [{\citenamefont {Dolev}\ \emph {et~al.}(2010)\citenamefont {Dolev},
  \citenamefont {Gross}, \citenamefont {Chung}, \citenamefont {Heiblum},
  \citenamefont {Umansky},\ and\ \citenamefont {Mahalu}}]{Dolev10}%
  \BibitemOpen
  \bibfield  {author} {\bibinfo {author} {\bibfnamefont {M.}~\bibnamefont
  {Dolev}}, \bibinfo {author} {\bibfnamefont {Y.}~\bibnamefont {Gross}},
  \bibinfo {author} {\bibfnamefont {Y.~C.}\ \bibnamefont {Chung}}, \bibinfo
  {author} {\bibfnamefont {M.}~\bibnamefont {Heiblum}}, \bibinfo {author}
  {\bibfnamefont {V.}~\bibnamefont {Umansky}},\ and\ \bibinfo {author}
  {\bibfnamefont {D.}~\bibnamefont {Mahalu}},\ }\href
  {https://doi.org/10.1103/PhysRevB.81.161303} {\bibfield  {journal} {\bibinfo
  {journal} {Phys. Rev. B}\ }\textbf {\bibinfo {volume} {81}},\ \bibinfo
  {pages} {161303} (\bibinfo {year} {2010})}\BibitemShut {NoStop}%
\bibitem [{\citenamefont {Kim}\ \emph {et~al.}(2024)\citenamefont {Kim},
  \citenamefont {Dev}, \citenamefont {Shaer}, \citenamefont {Kumar},
  \citenamefont {Ilin}, \citenamefont {Haug}, \citenamefont {Iskoz},
  \citenamefont {Watanabe}, \citenamefont {Taniguchi}, \citenamefont {Mross},
  \citenamefont {Stern},\ and\ \citenamefont {Ronen}}]{Ronen25}%
  \BibitemOpen
  \bibfield  {author} {\bibinfo {author} {\bibfnamefont {J.}~\bibnamefont
  {Kim}}, \bibinfo {author} {\bibfnamefont {H.}~\bibnamefont {Dev}}, \bibinfo
  {author} {\bibfnamefont {A.}~\bibnamefont {Shaer}}, \bibinfo {author}
  {\bibfnamefont {R.}~\bibnamefont {Kumar}}, \bibinfo {author} {\bibfnamefont
  {A.}~\bibnamefont {Ilin}}, \bibinfo {author} {\bibfnamefont {A.}~\bibnamefont
  {Haug}}, \bibinfo {author} {\bibfnamefont {S.}~\bibnamefont {Iskoz}},
  \bibinfo {author} {\bibfnamefont {K.}~\bibnamefont {Watanabe}}, \bibinfo
  {author} {\bibfnamefont {T.}~\bibnamefont {Taniguchi}}, \bibinfo {author}
  {\bibfnamefont {D.~F.}\ \bibnamefont {Mross}}, \bibinfo {author}
  {\bibfnamefont {A.}~\bibnamefont {Stern}},\ and\ \bibinfo {author}
  {\bibfnamefont {Y.}~\bibnamefont {Ronen}},\ }\href
  {https://doi.org/10.48550/ARXIV.2412.19886} {\bibinfo {title} {Aharonov-bohm
  interference in even-denominator fractional quantum hall states}} (\bibinfo
  {year} {2024})\BibitemShut {NoStop}%
\bibitem [{\citenamefont {Yutushui}\ \emph {et~al.}(2025)\citenamefont
  {Yutushui}, \citenamefont {Hermanns},\ and\ \citenamefont
  {Mross}}]{Yutushui25A}%
  \BibitemOpen
  \bibfield  {author} {\bibinfo {author} {\bibfnamefont {M.}~\bibnamefont
  {Yutushui}}, \bibinfo {author} {\bibfnamefont {M.}~\bibnamefont {Hermanns}},\
  and\ \bibinfo {author} {\bibfnamefont {D.~F.}\ \bibnamefont {Mross}},\ }\href
  {https://doi.org/10.48550/ARXIV.2502.12245} {\bibinfo {title} {Non-abelian
  phases from the condensation of abelian anyons}} (\bibinfo {year}
  {2025})\BibitemShut {NoStop}%
\bibitem [{\citenamefont {Nakamura}\ \emph {et~al.}(2020)\citenamefont
  {Nakamura}, \citenamefont {Liang}, \citenamefont {Gardner},\ and\
  \citenamefont {Manfra}}]{Nakamura20}%
  \BibitemOpen
  \bibfield  {author} {\bibinfo {author} {\bibfnamefont {J.}~\bibnamefont
  {Nakamura}}, \bibinfo {author} {\bibfnamefont {S.}~\bibnamefont {Liang}},
  \bibinfo {author} {\bibfnamefont {G.~C.}\ \bibnamefont {Gardner}},\ and\
  \bibinfo {author} {\bibfnamefont {M.~J.}\ \bibnamefont {Manfra}},\
  }\href@noop {} {\bibfield  {journal} {\bibinfo  {journal} {Nature Physics}\
  }\textbf {\bibinfo {volume} {16}},\ \bibinfo {pages} {931} (\bibinfo {year}
  {2020})}\BibitemShut {NoStop}%
\bibitem [{\citenamefont {Ghosh}\ \emph {et~al.}(2024)\citenamefont {Ghosh},
  \citenamefont {Labendik}, \citenamefont {Musina}, \citenamefont {Umansky},
  \citenamefont {Heiblum},\ and\ \citenamefont {Mross}}]{Ghosh24}%
  \BibitemOpen
  \bibfield  {author} {\bibinfo {author} {\bibfnamefont {B.}~\bibnamefont
  {Ghosh}}, \bibinfo {author} {\bibfnamefont {M.}~\bibnamefont {Labendik}},
  \bibinfo {author} {\bibfnamefont {L.}~\bibnamefont {Musina}}, \bibinfo
  {author} {\bibfnamefont {V.}~\bibnamefont {Umansky}}, \bibinfo {author}
  {\bibfnamefont {M.}~\bibnamefont {Heiblum}},\ and\ \bibinfo {author}
  {\bibfnamefont {D.~F.}\ \bibnamefont {Mross}}\ }\href
  {https://doi.org/10.48550/arXiv.2410.16488} {10.48550/arXiv.2410.16488}
  (\bibinfo {year} {2024})\BibitemShut {NoStop}%
\bibitem [{\citenamefont {Werkmeister}\ \emph {et~al.}(2025)\citenamefont
  {Werkmeister}, \citenamefont {Ehrets}, \citenamefont {Wesson}, \citenamefont
  {Najafabadi}, \citenamefont {Watanabe}, \citenamefont {Taniguchi},
  \citenamefont {Halperin}, \citenamefont {Yacoby},\ and\ \citenamefont
  {Kim}}]{Werkmeister25}%
  \BibitemOpen
  \bibfield  {author} {\bibinfo {author} {\bibfnamefont {T.}~\bibnamefont
  {Werkmeister}}, \bibinfo {author} {\bibfnamefont {J.~R.}\ \bibnamefont
  {Ehrets}}, \bibinfo {author} {\bibfnamefont {M.~E.}\ \bibnamefont {Wesson}},
  \bibinfo {author} {\bibfnamefont {D.~H.}\ \bibnamefont {Najafabadi}},
  \bibinfo {author} {\bibfnamefont {K.}~\bibnamefont {Watanabe}}, \bibinfo
  {author} {\bibfnamefont {T.}~\bibnamefont {Taniguchi}}, \bibinfo {author}
  {\bibfnamefont {B.~I.}\ \bibnamefont {Halperin}}, \bibinfo {author}
  {\bibfnamefont {A.}~\bibnamefont {Yacoby}},\ and\ \bibinfo {author}
  {\bibfnamefont {P.}~\bibnamefont {Kim}},\ }\bibfield  {journal} {\bibinfo
  {journal} {Science}\ }\href {https://doi.org/10.1126/science.adp5015}
  {10.1126/science.adp5015} (\bibinfo {year} {2025})\BibitemShut {NoStop}%
\bibitem [{\citenamefont {Samuelson}\ \emph {et~al.}(2024)\citenamefont
  {Samuelson}, \citenamefont {Cohen}, \citenamefont {Wang}, \citenamefont
  {Blanch}, \citenamefont {Taniguchi}, \citenamefont {Watanabe}, \citenamefont
  {Zaletel},\ and\ \citenamefont {Young}}]{Samuelson24}%
  \BibitemOpen
  \bibfield  {author} {\bibinfo {author} {\bibfnamefont {N.~L.}\ \bibnamefont
  {Samuelson}}, \bibinfo {author} {\bibfnamefont {L.~A.}\ \bibnamefont
  {Cohen}}, \bibinfo {author} {\bibfnamefont {W.}~\bibnamefont {Wang}},
  \bibinfo {author} {\bibfnamefont {S.}~\bibnamefont {Blanch}}, \bibinfo
  {author} {\bibfnamefont {T.}~\bibnamefont {Taniguchi}}, \bibinfo {author}
  {\bibfnamefont {K.}~\bibnamefont {Watanabe}}, \bibinfo {author}
  {\bibfnamefont {M.~P.}\ \bibnamefont {Zaletel}},\ and\ \bibinfo {author}
  {\bibfnamefont {A.~F.}\ \bibnamefont {Young}},\ }\href@noop {} {\bibfield
  {journal} {\bibinfo  {journal} {arXiv preprint arXiv:2403.19628}\ } (\bibinfo
  {year} {2024})}\BibitemShut {NoStop}%
\bibitem [{\citenamefont {Parameswaran}\ \emph {et~al.}(2011)\citenamefont
  {Parameswaran}, \citenamefont {Kivelson}, \citenamefont {Sondhi},\ and\
  \citenamefont {Spivak}}]{Parameswaran11}%
  \BibitemOpen
  \bibfield  {author} {\bibinfo {author} {\bibfnamefont {S.~A.}\ \bibnamefont
  {Parameswaran}}, \bibinfo {author} {\bibfnamefont {S.~A.}\ \bibnamefont
  {Kivelson}}, \bibinfo {author} {\bibfnamefont {S.~L.}\ \bibnamefont
  {Sondhi}},\ and\ \bibinfo {author} {\bibfnamefont {B.~Z.}\ \bibnamefont
  {Spivak}},\ }\href {https://doi.org/10.1103/PhysRevLett.106.236801}
  {\bibfield  {journal} {\bibinfo  {journal} {Phys. Rev. Lett.}\ }\textbf
  {\bibinfo {volume} {106}},\ \bibinfo {pages} {236801} (\bibinfo {year}
  {2011})}\BibitemShut {NoStop}%
\bibitem [{\citenamefont {Parameswaran}\ \emph {et~al.}(2012)\citenamefont
  {Parameswaran}, \citenamefont {Kivelson}, \citenamefont {Rezayi},
  \citenamefont {Simon}, \citenamefont {Sondhi},\ and\ \citenamefont
  {Spivak}}]{Parameswaran12}%
  \BibitemOpen
  \bibfield  {author} {\bibinfo {author} {\bibfnamefont {S.~A.}\ \bibnamefont
  {Parameswaran}}, \bibinfo {author} {\bibfnamefont {S.~A.}\ \bibnamefont
  {Kivelson}}, \bibinfo {author} {\bibfnamefont {E.~H.}\ \bibnamefont
  {Rezayi}}, \bibinfo {author} {\bibfnamefont {S.~H.}\ \bibnamefont {Simon}},
  \bibinfo {author} {\bibfnamefont {S.~L.}\ \bibnamefont {Sondhi}},\ and\
  \bibinfo {author} {\bibfnamefont {B.~Z.}\ \bibnamefont {Spivak}},\ }\href
  {https://doi.org/10.1103/PhysRevB.85.241307} {\bibfield  {journal} {\bibinfo
  {journal} {Phys. Rev. B}\ }\textbf {\bibinfo {volume} {85}},\ \bibinfo
  {pages} {241307} (\bibinfo {year} {2012})}\BibitemShut {NoStop}%
\end{thebibliography}%


\begin{thebibliography}{16}%
\makeatletter
\providecommand \@ifxundefined [1]{%
 \@ifx{#1\undefined}
}%
\providecommand \@ifnum [1]{%
 \ifnum #1\expandafter \@firstoftwo
 \else \expandafter \@secondoftwo
 \fi
}%
\providecommand \@ifx [1]{%
 \ifx #1\expandafter \@firstoftwo
 \else \expandafter \@secondoftwo
 \fi
}%
\providecommand \natexlab [1]{#1}%
\providecommand \enquote  [1]{``#1''}%
\providecommand \bibnamefont  [1]{#1}%
\providecommand \bibfnamefont [1]{#1}%
\providecommand \citenamefont [1]{#1}%
\providecommand \href@noop [0]{\@secondoftwo}%
\providecommand \href [0]{\begingroup \@sanitize@url \@href}%
\providecommand \@href[1]{\@@startlink{#1}\@@href}%
\providecommand \@@href[1]{\endgroup#1\@@endlink}%
\providecommand \@sanitize@url [0]{\catcode `\\12\catcode `\$12\catcode
  `\&12\catcode `\#12\catcode `\^12\catcode `\_12\catcode `\%12\relax}%
\providecommand \@@startlink[1]{}%
\providecommand \@@endlink[0]{}%
\providecommand \url  [0]{\begingroup\@sanitize@url \@url }%
\providecommand \@url [1]{\endgroup\@href {#1}{\urlprefix }}%
\providecommand \urlprefix  [0]{URL }%
\providecommand \Eprint [0]{\href }%
\providecommand \doibase [0]{https://doi.org/}%
\providecommand \selectlanguage [0]{\@gobble}%
\providecommand \bibinfo  [0]{\@secondoftwo}%
\providecommand \bibfield  [0]{\@secondoftwo}%
\providecommand \translation [1]{[#1]}%
\providecommand \BibitemOpen [0]{}%
\providecommand \bibitemStop [0]{}%
\providecommand \bibitemNoStop [0]{.\EOS\space}%
\providecommand \EOS [0]{\spacefactor3000\relax}%
\providecommand \BibitemShut  [1]{\csname bibitem#1\endcsname}%
\let\auto@bib@innerbib\@empty
\bibitem [{\citenamefont {Haldane}(1983)}]{Haldane83}%
  \BibitemOpen
  \bibfield  {author} {\bibinfo {author} {\bibfnamefont {F.~D.~M.}\
  \bibnamefont {Haldane}},\ }\href {https://doi.org/10.1103/PhysRevLett.51.605}
  {\bibfield  {journal} {\bibinfo  {journal} {Phys. Rev. Lett.}\ }\textbf
  {\bibinfo {volume} {51}},\ \bibinfo {pages} {605} (\bibinfo {year}
  {1983})}\BibitemShut {NoStop}%
\bibitem [{\citenamefont {Wu}\ and\ \citenamefont {Yang}(1976)}]{Wu76}%
  \BibitemOpen
  \bibfield  {author} {\bibinfo {author} {\bibfnamefont {T.~T.}\ \bibnamefont
  {Wu}}\ and\ \bibinfo {author} {\bibfnamefont {C.~N.}\ \bibnamefont {Yang}},\
  }\href {https://doi.org/10.1016/0550-3213(76)90143-7} {\bibfield  {journal}
  {\bibinfo  {journal} {Nuclear Physics B}\ }\textbf {\bibinfo {volume}
  {107}},\ \bibinfo {pages} {365–380} (\bibinfo {year} {1976})}\BibitemShut
  {NoStop}%
\bibitem [{\citenamefont {Wu}\ and\ \citenamefont {Yang}(1977)}]{Wu77}%
  \BibitemOpen
  \bibfield  {author} {\bibinfo {author} {\bibfnamefont {T.~T.}\ \bibnamefont
  {Wu}}\ and\ \bibinfo {author} {\bibfnamefont {C.~N.}\ \bibnamefont {Yang}},\
  }\href {https://doi.org/10.1103/physrevd.16.1018} {\bibfield  {journal}
  {\bibinfo  {journal} {Physical Review D}\ }\textbf {\bibinfo {volume} {16}},\
  \bibinfo {pages} {1018–1021} (\bibinfo {year} {1977})}\BibitemShut
  {NoStop}%
\bibitem [{\citenamefont {Jain}\ and\ \citenamefont
  {Kamilla}(1997{\natexlab{a}})}]{Jain97}%
  \BibitemOpen
  \bibfield  {author} {\bibinfo {author} {\bibfnamefont {J.~K.}\ \bibnamefont
  {Jain}}\ and\ \bibinfo {author} {\bibfnamefont {R.~K.}\ \bibnamefont
  {Kamilla}},\ }\href {https://doi.org/10.1142/S0217979297001301} {\bibfield
  {journal} {\bibinfo  {journal} {Int. J. Mod. Phys. B}\ }\textbf {\bibinfo
  {volume} {11}},\ \bibinfo {pages} {2621} (\bibinfo {year}
  {1997}{\natexlab{a}})}\BibitemShut {NoStop}%
\bibitem [{\citenamefont {Jain}\ and\ \citenamefont
  {Kamilla}(1997{\natexlab{b}})}]{Jain97B}%
  \BibitemOpen
  \bibfield  {author} {\bibinfo {author} {\bibfnamefont {J.~K.}\ \bibnamefont
  {Jain}}\ and\ \bibinfo {author} {\bibfnamefont {R.~K.}\ \bibnamefont
  {Kamilla}},\ }\href {https://doi.org/10.1103/PhysRevB.55.R4895} {\bibfield
  {journal} {\bibinfo  {journal} {Phys. Rev. B}\ }\textbf {\bibinfo {volume}
  {55}},\ \bibinfo {pages} {R4895} (\bibinfo {year}
  {1997}{\natexlab{b}})}\BibitemShut {NoStop}%
\bibitem [{\citenamefont {Jain}(2007)}]{Jain07}%
  \BibitemOpen
  \bibfield  {author} {\bibinfo {author} {\bibfnamefont {J.~K.}\ \bibnamefont
  {Jain}},\ }\href@noop {} {\emph {\bibinfo {title} {Composite Fermions}}}\
  (\bibinfo  {publisher} {Cambridge University Press, New York, US},\ \bibinfo
  {year} {2007})\BibitemShut {NoStop}%
\bibitem [{\citenamefont {Jain}(1989{\natexlab{a}})}]{Jain89}%
  \BibitemOpen
  \bibfield  {author} {\bibinfo {author} {\bibfnamefont {J.~K.}\ \bibnamefont
  {Jain}},\ }\href {https://doi.org/10.1103/PhysRevLett.63.199} {\bibfield
  {journal} {\bibinfo  {journal} {Phys. Rev. Lett.}\ }\textbf {\bibinfo
  {volume} {63}},\ \bibinfo {pages} {199} (\bibinfo {year}
  {1989}{\natexlab{a}})}\BibitemShut {NoStop}%
\bibitem [{\citenamefont {Gattu}\ and\ \citenamefont {Jain}(2025)}]{Gattu25}%
  \BibitemOpen
  \bibfield  {author} {\bibinfo {author} {\bibfnamefont {M.}~\bibnamefont
  {Gattu}}\ and\ \bibinfo {author} {\bibfnamefont {J.}~\bibnamefont {Jain}},\
  }\bibfield  {journal} {\bibinfo  {journal} {Physical Review Letters}\
  }\textbf {\bibinfo {volume} {134}},\ \href
  {https://doi.org/10.1103/physrevlett.134.156501}
  {10.1103/physrevlett.134.156501} (\bibinfo {year} {2025})\BibitemShut
  {NoStop}%
\bibitem [{\citenamefont {Balram}\ \emph {et~al.}(2013)\citenamefont {Balram},
  \citenamefont {W\'ojs},\ and\ \citenamefont {Jain}}]{Balram13}%
  \BibitemOpen
  \bibfield  {author} {\bibinfo {author} {\bibfnamefont {A.~C.}\ \bibnamefont
  {Balram}}, \bibinfo {author} {\bibfnamefont {A.}~\bibnamefont {W\'ojs}},\
  and\ \bibinfo {author} {\bibfnamefont {J.~K.}\ \bibnamefont {Jain}},\ }\href
  {https://doi.org/10.1103/PhysRevB.88.205312} {\bibfield  {journal} {\bibinfo
  {journal} {Phys. Rev. B}\ }\textbf {\bibinfo {volume} {88}},\ \bibinfo
  {pages} {205312} (\bibinfo {year} {2013})}\BibitemShut {NoStop}%
\bibitem [{\citenamefont {Mandal}\ and\ \citenamefont {Jain}(2002)}]{Mandal02}%
  \BibitemOpen
  \bibfield  {author} {\bibinfo {author} {\bibfnamefont {S.~S.}\ \bibnamefont
  {Mandal}}\ and\ \bibinfo {author} {\bibfnamefont {J.~K.}\ \bibnamefont
  {Jain}},\ }\href {https://doi.org/10.1103/PhysRevB.66.155302} {\bibfield
  {journal} {\bibinfo  {journal} {Phys. Rev. B}\ }\textbf {\bibinfo {volume}
  {66}},\ \bibinfo {pages} {155302} (\bibinfo {year} {2002})}\BibitemShut
  {NoStop}%
\bibitem [{\citenamefont {Kamilla}\ \emph {et~al.}(1996)\citenamefont
  {Kamilla}, \citenamefont {Wu},\ and\ \citenamefont {Jain}}]{Kamilla96}%
  \BibitemOpen
  \bibfield  {author} {\bibinfo {author} {\bibfnamefont {R.~K.}\ \bibnamefont
  {Kamilla}}, \bibinfo {author} {\bibfnamefont {X.~G.}\ \bibnamefont {Wu}},\
  and\ \bibinfo {author} {\bibfnamefont {J.~K.}\ \bibnamefont {Jain}},\
  }\href@noop {} {\bibfield  {journal} {\bibinfo  {journal} {Solid State
  Commun.}\ }\textbf {\bibinfo {volume} {99}} (\bibinfo {year}
  {1996})}\BibitemShut {NoStop}%
\bibitem [{\citenamefont {Ortalano}\ \emph {et~al.}(1997)\citenamefont
  {Ortalano}, \citenamefont {He},\ and\ \citenamefont
  {Das~Sarma}}]{Ortalano97}%
  \BibitemOpen
  \bibfield  {author} {\bibinfo {author} {\bibfnamefont {M.~W.}\ \bibnamefont
  {Ortalano}}, \bibinfo {author} {\bibfnamefont {S.}~\bibnamefont {He}},\ and\
  \bibinfo {author} {\bibfnamefont {S.}~\bibnamefont {Das~Sarma}},\ }\href
  {https://doi.org/10.1103/PhysRevB.55.7702} {\bibfield  {journal} {\bibinfo
  {journal} {Phys. Rev. B}\ }\textbf {\bibinfo {volume} {55}},\ \bibinfo
  {pages} {7702} (\bibinfo {year} {1997})}\BibitemShut {NoStop}%
\bibitem [{\citenamefont {Rother}(2020)}]{Martin20}%
  \BibitemOpen
  \bibfield  {author} {\bibinfo {author} {\bibfnamefont {M.}~\bibnamefont
  {Rother}},\ }\href@noop {} {\bibinfo {title} {2{D} {Schroedinger} {Poisson}
  solver {A}{Q}{I}{L}{A}}},\ \bibinfo {howpublished}
  {\url{https://www.mathworks.com/matlabcentral/fileexchange/3344-2d-schroedinger-poisson-solver-aquila}}
  (\bibinfo {year} {2009--2020})\BibitemShut {NoStop}%
\bibitem [{\citenamefont {Balram}\ \emph {et~al.}(2018)\citenamefont {Balram},
  \citenamefont {Barkeshli},\ and\ \citenamefont {Rudner}}]{Balram18}%
  \BibitemOpen
  \bibfield  {author} {\bibinfo {author} {\bibfnamefont {A.~C.}\ \bibnamefont
  {Balram}}, \bibinfo {author} {\bibfnamefont {M.}~\bibnamefont {Barkeshli}},\
  and\ \bibinfo {author} {\bibfnamefont {M.~S.}\ \bibnamefont {Rudner}},\
  }\href {https://doi.org/10.1103/PhysRevB.98.035127} {\bibfield  {journal}
  {\bibinfo  {journal} {Phys. Rev. B}\ }\textbf {\bibinfo {volume} {98}},\
  \bibinfo {pages} {035127} (\bibinfo {year} {2018})}\BibitemShut {NoStop}%
\bibitem [{\citenamefont {Jain}(1989{\natexlab{b}})}]{Jain89B}%
  \BibitemOpen
  \bibfield  {author} {\bibinfo {author} {\bibfnamefont {J.~K.}\ \bibnamefont
  {Jain}},\ }\href {https://doi.org/10.1103/PhysRevB.40.8079} {\bibfield
  {journal} {\bibinfo  {journal} {Phys. Rev. B}\ }\textbf {\bibinfo {volume}
  {40}},\ \bibinfo {pages} {8079} (\bibinfo {year}
  {1989}{\natexlab{b}})}\BibitemShut {NoStop}%
\bibitem [{\citenamefont {Yutushui}\ and\ \citenamefont
  {Mross}(2024)}]{Yutushui24}%
  \BibitemOpen
  \bibfield  {author} {\bibinfo {author} {\bibfnamefont {M.}~\bibnamefont
  {Yutushui}}\ and\ \bibinfo {author} {\bibfnamefont {D.~F.}\ \bibnamefont
  {Mross}},\ }\href {http://dx.doi.org/10.1103/PhysRevB.92.235302} {\bibfield
  {journal} {\bibinfo  {journal} {arXiv [cond-mat.str-el]}\ } (\bibinfo {year}
  {2024})}\BibitemShut {NoStop}%
\end{thebibliography}%

{\bf Acknowledgments.} We are grateful to Moty Heiblum for sharing with us unpublished results that prompted our study, and him, Dmitri Feldman and Steven Kivelson for valuable discussions and feedback on our work. The work was supported in part by the U. S. Department of Energy, Office of Basic Energy Sciences, under Grant no. DE-SC0005042. 

{\bf Author Contributions.} MG performed the calculations. MG and JKJ conceived and planned the project and wrote the paper.

{\bf Competing Interests.} The authors declare no competing interests.


\end{document}


\title{Supplementary Information for\\ ``Molecular anyons in the fractional quantum Hall effect"}
\author{Mytraya Gattu~\orcidlink{0000-0001-6994-389X} and J. K. Jain~\orcidlink{000-0003-0082-5881}}
\affiliation{Department of Physics, 104 Davey Lab, Pennsylvania State University, University Park, Pennsylvania 16802,USA}
\date{\today}
\maketitle



In this Supplementary Information, we provide the technical details underlying our study of molecular anyon formation in fractional quantum Hall (FQH) systems.  In Section \ref{sec:spherical} we review the spherical geometry, used for our calculations. This follows the formulation of composite-fermion (CF) theory in Sec.~\ref{sec:CFwf}, including the construction of incompressible Jain states at $\nu=n/(2pn+1)$ and their lowest-energy quasiparticle (QP) and quasihole (QH) excitations.  Section~\ref{sec:MA} describes our two-stage methodology—CF diagonalization of multi-QP/QH basis states followed by binding-energy $\Delta_{q}$ calculations—to identify and characterize stable QP and QH molecules. Finally, Section~\ref{sec:parton} introduces the $\bar{2}\bar{2}111$ parton ansatz for the anti-Pfaffian phase at $\nu=5/2$ and presents the explicit wavefunctions for its non-topological and topological two-QH states. 

\section{Spherical geometry}
\label{sec:spherical}

We first introduce the spherical geometry~\cite{Haldane83}, used to study the bulk of fractional quantum Hall (FQH) liquids. In this  geometry, electrons move on a spherical surface $\mathcal{S}$. The surface $\mathcal{S}$ encloses a magnetic monopole of strength $Q$ that produces a magnetic field $\mathbf{B}(\mathbf{r}) = 2Q\phi_{0}/(4\pi r^{3})\bm{r}$, which is uniform and normal to the surface $\mathcal{S}$ everywhere, producing a total flux $2Q\phi_{0}$. Here, $\phi_{0} = hc / e$ is the flux quantum and $2Q$ is an integer. In terms of the magnetic length, $\ell = \sqrt{\hbar c/eB}$, where $B$ is the strength of the magnetic field on $\mathcal{S}$, and the radius is given by $R = \sqrt{Q}\ell$.

The single-particle eigenfunctions on $\mathcal{S}$ are the monopole harmonics~\cite{Wu76,Wu77} $Y_{Q, l, m}(\theta, \phi)$, which are eigenstates of the orbital $\hat{L}$ and azimuthal $\hat{L}_{z}$ angular momentum operators with quantum numbers $l = \abs{Q} + n_{l}$ and $m = -l, \dots, l$ respectively. Here, $\theta$ and $\phi$ are the polar and azimuthal angles describing the position of an electron on $\mathcal{S}$ and $n_{\rm LL}=0, 1, \dots$ is the Landau Level (LL) index. The degeneracy of a LL is $2l+1$. 


A convenient representation of the monopole harmonics is in terms of the spinor variables~\cite{Haldane83} $u = \cos(\theta/2)e^{i\phi/2}$ and $v = \sin(\theta/2)e^{-i\phi/2}$. In terms of $u$ and $v$, $Y_{Q, l, m}(\theta, \phi)$ can be expressed as~\cite{Jain97,Jain97B,Jain07}:
\begin{equation}
  \begin{split}
    &Y_{Q, l, m}(\theta, \phi) = N_{Q, l, m}(-1)^{l-m}v^{Q-m}u^{Q+m} \times \\
    &\sum_{s=0}^{l-m}(-1)^{s}\tbinom{l-Q}{s}\tbinom{l+Q}{l-m-s}(v^{\star}v)^{l-Q-s}(u^{\star}u)^{s} \\
    &N_{Q, l, m} = \sqrt{\tfrac{(2l+1) (l-m)! (l+m)!}{4\pi (l-Q)! (l+Q)!}}
  \end{split}
\end{equation}
In the lowest LL ($n_{\rm LL}=0, l=Q$), this simplifies to $Y_{Q, Q, m}(\theta, \phi) = N_{Q, Q, m}(-1)^{Q-m}v^{Q-m}u^{Q+m}$.

\section{Composite fermion (CF) wavefunctions}
\label{sec:CFwf}

Having described the orbitals on $\mathcal{S}$, we now outline the construction of the many-body CF wavefunctions for the incompressible states and their lowest energy QP and QH excitations at Jain fillings $\nu = n/(2pn+1)$.

Consider a system of $N$ electrons confined to the LLL, each electron binding $2p$ vortices to form CFs at effective monopole strength $Q^*=Q-p(N-1)$. When these CFs fill $n$ $\Lambda$Ls, the resulting state at electron filling $\nu=n/(2pn+1)$ is incompressible and can be written as~\cite{Jain89, Jain07}
\begin{equation}\label{eq:jain-incompressible}
 \begin{split}
  \Psi_{\frac{n}{2pn+1}}(\Omega_{1}, \dots, \Omega_{N}) = \LLL \Phi_{n}(\Omega_{1}, \dots, \Omega_{N}; Q^{\star})\Phi_{1}^{2p}
 \end{split}
\end{equation}
Here, $\LLL$ is the projection operator into the LLL, $\Omega_{i} \equiv (\theta_{i}, \phi_{i})$ denotes the $i^{\mathrm{th}}$ electron's position on $\mathcal{S}$, $\Phi_{n}(\Omega_{1}, \dots, \Omega_{N}; Q^{\star})$ is the Slater determinant of $N$ electrons at monopole strength $Q^{\star} = (N/n-n)/2$ filling $n$ LLs, and $\Phi_{1}^{2p}(\Omega_{1}, \dots, \Omega_{N}) = \prod_{i<j=1}^{N}(u_{i}v_{j}-u_{j}v_{i})^{2p}$ is the Jastrow factor (with each $\Phi_{1}$ factor at monopole strength $Q_{1} = (N-1)/2$-corresponding to the $\nu=1$ state of $N$ electrons), which attaches $2p$ vortices to each electron.

The single QP state at $\nu=n/(2pn+1)$ is obtained by adding one CF to the $n^{\rm th}$ $\Lambda$L: on top of the $n$ filled $\Lambda$Ls. It carries total angular momentum $L_{\rm QP}=N/2+(n-1/n)/2$ and, in the $L_z = m$ sector, its wave function is
\begin{equation}\label{eq:cf-qp}
  \begin{split}
    \Psi_{\frac{n}{2pn+1}}^{\mathrm{QP};m}(\Omega_{1}, \dots, \Omega_{N}) = \LLL \Phi_{n}^{+; m}(\Omega_{1}, \dots, \Omega_{N}; Q^{\star}) \Phi_{1}^{2p}
  \end{split}
\end{equation}
Here, $\Phi_{n}^{+; m}$ is a Slater determinant of $N$ electrons at monopole strength $Q^{\star} = N/2n-(n+1/n)/2$, with the first $n$ LLs filled and one additional electron in the $L_{z} = m$ orbital in the $n^{\rm th}$ LL.

The single QH state at $\nu = n/(2pn+1)$, formed by removing a CF from the $(n-1)^{\rm th}$ $\Lambda$L, occurs at $L_{\rm QH}=N/2n+(n+1/n)/2-1$ and, for $L_{z}=m$ has the wavefunction:
\begin{equation}\label{eq:cf-qh}
  \begin{split}
    \Psi_{\frac{n}{2pn+1}}^{\mathrm{QH};m}(\Omega_{1}, \dots, \Omega_{N}) = \LLL \Phi_{n}^{-;m}(\Omega_{1}, \dots, \Omega_{N}; Q^{\star}) \Phi_{1}^{2p}
  \end{split}
\end{equation}
Here, $\Phi_{n}^{-; m}$ is a Slater determinant of $N$ electrons at monopole strength $Q^{\star} = N/2n-(n-1/n)/2$, with all orbitals in the first $n$ LLs filled except for the $L_{z}=-m$ orbital in the $(n-1)^{\rm th}$ LL.

In our search for the lowest energy molecular anyons, we consider only the subspace of CFs with the lowest ``CF kinetic energy," i.e., only those configurations in which all QHs are in the topmost occupied ($n-1^{\rm th}$) $\Lambda$L and all QPs are in the lowest unoccupied ($n^{\rm th}$) $\Lambda$L. One may ask if higher CF kinetic energy configurations might also play a role. These would involve a neutral QP-QH excitation, i.e., a CF exciton. 
The lowest‐energy CF exciton is formed by promoting one CF from the $(n-1)$th $\Lambda$L into the $n$th. Its excitation energy—$\sim0.10\,e^2/\varepsilon\ell$ at $\nu=1/3$, $\sim0.06\,e^2/\varepsilon\ell$ at $\nu=2/5$, and $\sim0.04\,e^2/\varepsilon\ell$ at $\nu=3/7$~\cite{Gattu25}—is an order of magnitude larger than the typical QP/QH binding energies ($\sim 0.002-0.005\,e^2/\varepsilon\ell$). Therefore, one might expect that these are not relevant to the issue of our interest. Explicit calculations have confirmed (e.g., see Ref.~\cite{Balram13}) that the subspace of the lowest CF kinetic energy provides an excellent account of the ED energies for states containing several QPs or QHs, showing that dressing by CF excitons causes negligible corrections to the energies of the molecular anyons. Consequently, we neglect CF excitons in our study.

Constructing multi–QP/QH states is more involved. Single‐QP/QH states occur at a fixed total $L$ (for fixed $N$), but multi‐QP / QH states populate multiple $L$ sectors according to the rules of angular momentum addition. One must therefore specify both $(L,L_z)$ when building these wave functions. Moreover, for three or more QPs/QHs (at intermediate $L$), fixing $(L,L_z)$ does not completely specify a state, and an additional label is required. We will use the Greek symbols $\alpha, \beta, \dots$ for this purpose.

The $\alpha^{\rm th}$ state containing $q$ QPs at the total $L=l, L_{z}=m$ can be written as:
\begin{equation}\label{eq:cf-multi-qps}
  \begin{split}
    &\Psi_{\frac{n}{2pn+1}}^{\mathrm{QP}; q, l, m, \alpha}(\Omega_{1}, \dots, \Omega_{N}) = \\
    &\LLL \Phi_{n}^{+; q, l, m}(\Omega_{1}, \dots, \Omega_{N}; Q^{\star}) \Phi_{1}^{2p}
  \end{split}
\end{equation}
Here, $\Phi_{n}^{+; q, l, m, \alpha}(\Omega_{1}, \dots, \Omega_{N}; Q^{\star})$ is the $\alpha^{\rm th}$ Slater determinant of $N$ electrons at monopole strength $Q^{\star} = (N/n-n)/2 - q/2n$, with the first $n$ LLs filled and $q$ additional electrons in the $n^{\rm th}$ LL, projected (employing Clebsch-Gordon coefficients) into the $L=l, L_{z}=m$ sector. Since $\LLL$ can annihilate some linear combinations, the number of nonzero $\Psi$ states never exceeds the number of $\phi$ basis states; labeling by $\alpha,\beta,\dots$ therefore refers back to the corresponding unprojected $\Phi_{n}^{+;q,l,m,\alpha}\Phi_{1}^{2p}$ states.  

Similarly, the $\alpha^{\rm th}$ state containing $q$ QHs at the $L=l, L_{z}=m$ can be written as:
\begin{equation}\label{eq:cf-multi-qhs}
  \begin{split}
    &\Psi_{\frac{n}{2pn+1}}^{\mathrm{QH}; q, l, m, \alpha}(\Omega_{1}, \dots, \Omega_{N}) = \\
    &\LLL \Phi_{n}^{-; q, l, m, \alpha}(\Omega_{1}, \dots, \Omega_{N}; Q^{\star})\Phi_{1}^{2p}
  \end{split}
\end{equation}
Here, $\Phi_{n}^{-; q, l, m, \alpha}(\Omega_{1}, \dots, \Omega_{N}; Q^{\star})$ is the $\alpha^{\rm th}$ Slater determinant of $N$ electrons at monopole strength $Q^{\star} = (N/n-n)/2 + q/2n$, with the first $n$ LLs filled by electrons completely except for $q$ orbitals in the $(n-1)^{\rm th}$ LL projected into the $L=l, L_{z}=m$ sector.

In the next section, we will use the above wave functions to identify and characterize stable QP and QH molecules.

\section{Identifying stable molecular anyons}
\label{sec:MA}

In this section, we detail our two-step approach to probing the formation of molecular anyons composed of QPs and QHs.  
First, for a fixed number $q$ of QPs or QHs, we perform CF diagonalization (CFD)~\cite{Mandal02} of the multi‐QP/QH states (Eqs.~\ref{eq:cf-multi-qps}, \ref{eq:cf-multi-qhs}) at moderate system sizes to identify the lowest‐energy configuration. Second, whenever a bound molecular anyon emerges, we calculate its binding energy $\Delta_q$ using the procedure summarized in the main text and described in detail below.

Armed with this road map, we now turn to the first stage.  To identify the ground‐state configuration of $q$ QPs/QHs at each total angular momentum $L=l$, we diagonalize the Coulomb interaction
$$
V_c \;=\;\sum_{i<j}\frac{1}{\sqrt{Q}\,\lvert\Omega_i-\Omega_j\rvert\,\ell}
$$
in the non-orthogonal CF basis $\{\Psi_{n/(2pn+1)}^{\mathrm{QP}/\mathrm{QH};q,l,\alpha}\}_{\alpha=1}$ (the $m$ label is dropped). To do so, we first compute the overlap matrix 
$$
O_{\alpha\beta} = \big\langle \Psi^{\QP{}/\QH{}; q, l,\alpha}\big|\Psi^{\QP{}/\QH{}; q, l,\beta}\big\rangle
$$
via Metropolis–Hastings–Gibbs sampling, diagonalize $O$ to eliminate null modes, and project the Coulomb matrix (also calculated using Metropolis–Hastings–Gibbs sampling) 
$$
V_{\alpha\beta} = \big\langle \Psi^{\QP{}/\QH{}; q, l,\alpha}\big|V_{c}\big|\Psi^{\QP{}/\QH{}; q, l,\beta}\big\rangle
$$
into the resulting orthonormal subspace. Diagonalizing this transformed interaction then yields the ground state.  All LLL projections use the Jain–Kamilla method~\cite{Kamilla96,Gattu25}.

\begin{figure*}
  \includegraphics[width=\figwidth]{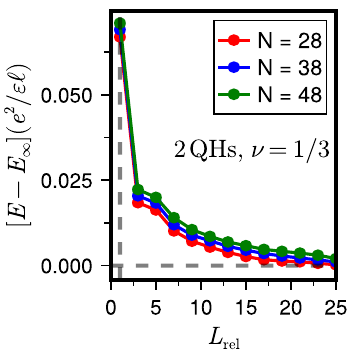}
  \includegraphics[width=\figwidth]{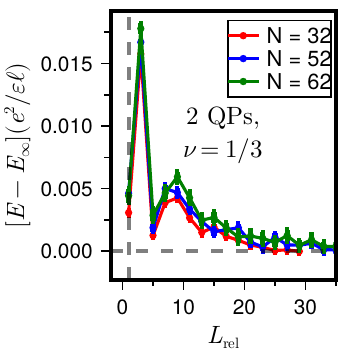}
  \includegraphics[width=\figwidth]{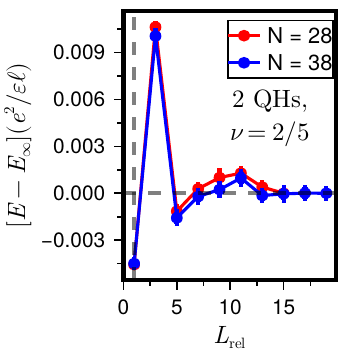}
  \includegraphics[width=\figwidth]{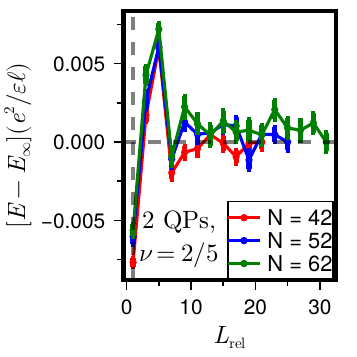}
  \includegraphics[width=\figwidth]{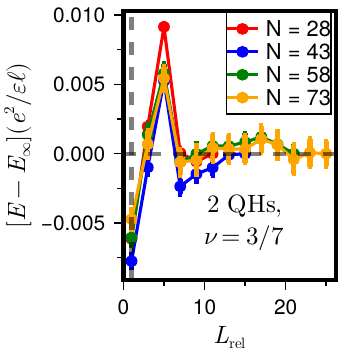}
  \includegraphics[width=\figwidth]{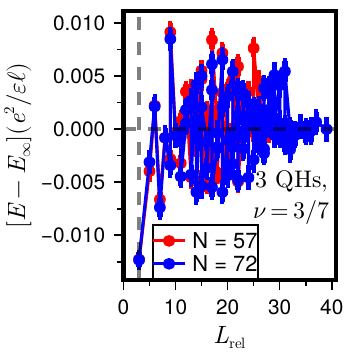}
  \includegraphics[width=\figwidth]{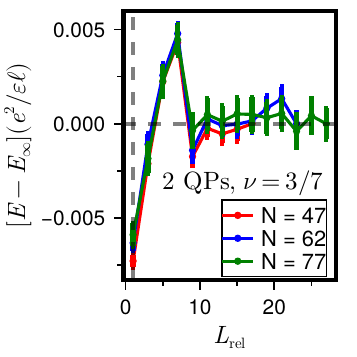}
  \includegraphics[width=\figwidth]{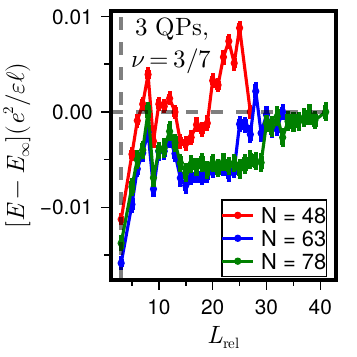}
  \includegraphics[width=\figwidth]{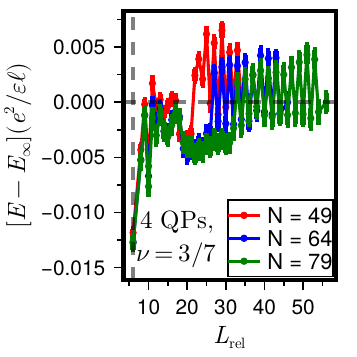}
  \caption{Energy $E$ (in units of $e^{2}/\varepsilon \ell$) of multi- QPs and QHs states at Jain fillings $\nu = 1/3, 2/5$ and $3/7$ as a function of their relative angular momentum $L_{\mathrm{rel}}$ for various system sizes $N$. All curves are shifted by $E_{\infty}$, the energy at the maximum $L_{\mathrm{rel}}$ for each $N$.}
  
    \label{fig:qps-qhs-dispersion}
\end{figure*}

Figure~\ref{fig:qps-qhs-dispersion} shows the energy of $q$-QP/QH states at $\nu=1/3$, $2/5$, and $3/7$ as a function of the relative angular momentum
$$
L_{\rm rel} \;=\; L_{\rm max} \;-\; L,
$$
which is the angular momentum measured relative to the maximum total angular momentum $L_{\rm max}$. For $q$ QPs or $q$ QHs, it is given by $L_{\rm max}=qL_{\rm QP}$ or $L_{\rm max}=qL_{\rm QH}$
with
$$
L_{\rm QP} = \frac{N}{2n} - \frac{q}{2n} + \frac{n}{2}, 
\quad
L_{\rm QH} = \frac{N}{2n} + \frac{n}{2} + \frac{q}{2n} - 1.
$$
It is noted that the smallest value of $L_{\rm rel}$ corresponds to the most compact configuration of the QPs or QHs, and the QPs/QHs are farther away as  $L_{\rm rel}$ increases. When the QPs/QHs are all far separated from one another, the interaction between them is negligible, and the energy is very close to $q$ times the energy of a single QP/QH, producing a continuum of states at this energy. 

A bound state exists whenever a level at finite $L_{\rm rel}$ falls below the asymptotic continuum. For the Jain fractions, we find that the lowest energy bound state always appears at the smallest allowed $L_{\rm rel}$, corresponding to the most compact arrangement of the $q$ excitations.  
For QPs, this is realized by placing $q$ CFs in the $n$th $\Lambda$L orbitals
$$
L_z = L_{\rm QP},\,L_{\rm QP}-1,\,\dots,\,L_{\rm QP}-q+1,
$$
side-by-side.  For QHs, it corresponds to placing $q$ holes in the $(n-1)$th $\Lambda$L orbitals
$$
L_z = -L_{\rm QH},\,-L_{\rm QH}+1,\,\dots,\,-L_{\rm QH}+q-1,
$$
side-by-side.  Having identified these lowest-energy QP$_q$ and QH$_q$ molecules, we now proceed to the second stage and determine how many QPs or QHs will bind to produce the lowest energy molecule.

For brevity we restrict the discussion to QPs; the QHs case is  analogous. We define the binding energy $\Delta_q$ of a $\mathrm{QP}_q$ molecule at $\nu=n/(2pn+1)$ by
$$
  \Delta_q = E_q^0 - \bigl(E_{q-1}^0 + E_1^0\bigr),
$$
where $E_q^0$ is the ground‐state energy of $\mathrm{QP}_q$, $E_{q-1}^0$ that of $\mathrm{QP}_{q-1}$, and $E_1^0$ that of a single isolated QP i.e. $\QP{1}$. The binding energy is thus the energy gain when we bring a QP from infinitely far and add it to $\mathrm{QP}_{q-1}$. 

Directly comparing these energies is unphysical, since they lie in different monopole sectors (and thus belong to different Hilbert spaces). Instead, we compute $\Delta_q$ at a fixed monopole strength by evaluating the energies of two CF configurations via Metropolis–Hastings–Gibbs sampling:
\begin{enumerate}
  \item $\mathrm{QP}_{q}$: Fill the first $n$ $\Lambda$Ls and place $q$ CFs in adjacent $L_z$ orbitals 
  $$
    L_z = L_{\mathrm{QP}},\,L_{\mathrm{QP}}-1,\,\dots,\,L_{\mathrm{QP}}-q+1
  $$
  of the $n^{\rm th}$ $\Lambda$L (localizing the $\mathrm{QP}_q$ at the North Pole). The energy of this configuration corresponds to $E^{0}_{q}$.
  \item $\QP{q-1}\otimes \QP{}$: Fill the first $n$ $\Lambda$Ls, place $q-1$ CFs in the orbitals
  $$
    L_z = L_{\mathrm{QP}},\,L_{\mathrm{QP}}-1,\,\dots,\,L_{\mathrm{QP}}-q+2
  $$
  of the $n^{\rm th}$ $\Lambda$L (localizing $\mathrm{QP}_{q-1}$ at the North Pole), and place one additional CF in the $L_z=-L_{\mathrm{QP}}$ orbital of the $n$th $\Lambda$L (localizing a single QP at the South Pole). The energy of this state, $E^0_{q-1,1}$, also contains a contribution coming from the interaction between the $q-1$ QPs at the North Pole and one at the South Pole. Correcting for that, we have $$E_{q-1}^0 + E_1^0=E^0_{q-1,1}-\frac{(q-1)e^{*2}}{2R},$$ where the last term is the interaction energy of the $q-1$ QPs at the North Pole and one at the South, with $e^*= 1/(2n+1)$. 
\end{enumerate}

Since both configurations occur at the same monopole strength, their energy difference directly yields $\Delta_{q}$. Beginning at $q=2$, we increase $q$ until $\Delta_{q}$ becomes positive i.e. an additional QP no longer binds. Figure 2 of the main text shows $\Delta_{q}$ computed as above, for the Coulomb interaction, for the QPs and QHs at $\nu = 2/5$ and $\nu=3/7$.

As discussed in the main text, the binding energy $\Delta_{q}$ depends sensitively on the electron–electron interaction. To capture deviations from the ideal 2D Coulomb case, we study the effects of finite quantum‐well width $w$ and transverse density $\rho$. Assuming that the lowest‐energy QP/QH configurations remain the most compact ones from the ideal 2D case, we employ a zero‐field local‐density approximation (LDA) to compute the transverse sub-band wave function and derive an effective 2D interaction~\cite{Ortalano97,Martin20} as a function of $w$ and $\rho$.  Figure~3 of the main text shows that increasing either $w$ or $\rho$  progressively unbinds larger QP/QH clusters, so that only smaller molecular anyons remain stable at $\nu = 2/5$ and $\nu = 3/7$.

\section{The $\bar{2}\bar{2}111$ parton model}
\label{sec:parton}

In this section, we first review the construction of the incompressible $\nu=5/2$ ground state in the spherical geometry using the $\bar{2}\bar{2}111$ parton model~\cite{Balram18}, and then describe how to build two QH states in the non‐topological (even $N$) and topological (odd $N$) sectors.

In the spherical geometry, the incompressible $\bar{2}\bar{2}111$ state of $N$ electrons, is written as
\begin{equation}\label{eq:22111-incompressible}
    \Psi_{1/2}(\Omega_{1}, \dots, \Omega_{N}) = \LLL [\conj{\Phi}_{2}(\Omega_{1}, \dots, \Omega_{N};Q^{\star})]^{2}\Phi_{1}^{3}
\end{equation}
Here, $\Phi_{2}$ is a Slater determinant of $N$ electrons at the monopole strength $Q^{\star}=N/4-1$ and $\Psi_{1/2}$ is at monopole strength $Q = N+1/2$. In practice, this state (and other parton states~\cite{Jain89B}) is evaluated in numerical calculations by writing it as 
\begin{equation}
    \Psi_{1/2}(\Omega_{1}, \dots, \Omega_{N}) = [\Psi_{2/3}]^2/\Phi_1
\end{equation}
where $\Psi_{2/3}=\LLL \conj{\Phi}_{2}\Phi_{1}^{2}$ is projected into the LLL using the Jain-Kamilla approach (the $\Phi_{1}$ factor requires no projection)~\cite{Jain97,Jain97B,Gattu25}.

A single QP or QH for the above state cannot be created on its own, because inserting one excitation into a single $\Phi_{2}$ yields an odd electron count in the other $\Phi_{2}$, which is forbidden for two filled levels on the sphere. The minimal charged excitations are therefore pairs of QPs or QHs.  We focus on two‐QH states.
Two distinct cases arise:
\begin{enumerate}
  \item \textbf{Non‐topological (even $N$).}  
    Both extra electrons occupy the same $\Phi_{2}$ factor. At $L=l,L_{z}=m$,
    \begin{equation}
    \begin{split}
        &\Psi_{1/2}^{\mathrm{non-topological}; l, m}(\Omega_{1}, \dots, \Omega_{N}) =\\
        &\LLL \conj{\Phi}_{2}{}^{+; 2, l, m}(\Omega_{1}, \dots, \Omega_{N}; Q^{\star}_{1})\conj{\Phi}_{2}(\Omega_{1}, \dots, \Omega_{N}; Q^{\star}_{2}) \Phi_{1}^{3}
    \end{split}        
    \end{equation}
    Here, $Q^{\star}_{1} = (N-6)/4$, $Q^{\star}_{2} = (N-4)/4$ and $L_{\mathrm{QH}} = (N+2)/4$. Here, the relative angular momentum $L_{\mathrm{rel}}=2L_{\mathrm{QH}}-l$ is odd i.e $L_{\mathrm{rel}}=1, 3, \dots$.
  \item \textbf{Topological (odd $N$).}  
    One extra electron is placed in each $\Phi_{2}$ factor and together projected into total angular momentum $L=l, L_{z}=m$ via Clebsch–Gordon coefficients $\overlap{l_{1}, m_{1}; l_{2}, m_{2}}{l, m}$ as:
    \begin{equation}
    \begin{split}
        &\Psi_{1/2}^{\mathrm{topological}; l, m}(\Omega_{1}, \dots, \Omega_{N}) =\\
        &\LLL \sum_{m_{1}}\overlap{L_{\mathrm{QH}},m_{1};L_{\mathrm{QH}},-m-m_{1}}{l,-m}\times \\
        &\conj{\Phi}_{2}{}^{+; m_{1}}(\Omega_{1}, \dots, \Omega_{N}; Q^{\star})\conj{\Phi}_{2}{}^{+;-m-m_{1}}(\Omega_{1}, \dots, \Omega_{N}; Q^{\star}) \Phi_{1}^{3}
    \end{split}        
    \end{equation}
    Here, $Q_1^{\star}=Q_2^{\star} = (N-5)/4$ and $L_{\mathrm{QH}} = (N+3)/4$. The relative angular momentum $L_{\mathrm{rel}}=2L_{\mathrm{QH}}-l$ is even, i.e. $L_{\mathrm{rel}}=0, 2, \dots$.
\end{enumerate}

We have used the above wavefunctions to evaluate the energy of two QH states (using the Coulomb interaction projected into the second LL~\cite{Yutushui24}) as a function of relative angular momentum $L_{\mathrm{rel}}$ using the Metropolis-Hastings-Gibbs algorithm. In Fig.~(4) of the main text, we have used these energies to extract the binding energy $\Delta_{2}$. In doing so, we have approximated the energy of two isolated QHs as the energy of the $L_{\mathrm{rel}} \rightarrow \infty$ state for two different system sizes $N$. We find that unlike for the Jain fractions, the lowest energy bound states appear at $L_{\mathrm{rel}}=5$ (non-topological) and $L_{\mathrm{rel}}=6$ (topological), i.e. they do not correspond to the most compact configurations. 
Given that the wave functions are not very accurate representations of the actual Coulomb wave functions, we have not pursued finding the most stable molecule of QHs. 

\bibliographystyle{apsrev4-2}
\bibliography{biblio_fqhe_processed.bib}